\documentclass[%
 reprint,
 amsmath,amssymb,
pre,
noeprint
]{revtex4-2}

\usepackage{bbold}
\usepackage{graphicx}% Include figure files
\usepackage{dcolumn}% Align table columns on decimal point
\usepackage{bm}% bold math
\usepackage[colorlinks=true,linkcolor=blue,citecolor=blue,urlcolor=blue]{hyperref}% add hypertext capabilities
\renewcommand{\vec}{\mathbf}

\newcommand{\beginsupplement}{%
        \setcounter{table}{0}
        \renewcommand{\thetable}{S\arabic{table}}%
        \setcounter{figure}{0}
        \renewcommand{\thefigure}{S\arabic{figure}}%
        \setcounter{equation}{0}
        \def\theequation{S\arabic{equation}}%
     }

\begin{document}

\preprint{APS/123-QED}

\title{Enhancing synchronization by optimal correlated noise}

\author{Sherwood Martineau}%
\author{Tim Saffold}%
\author{Timothy T. Chang}
\author{Henrik Ronellenfitsch}
\email{henrik.ronellenfitsch@gmail.com}
\affiliation{%
Physics Department, Williams College, 33 Lab Campus Drive, Williamstown, MA 01267, U.S.A.
}%

\date{\today}% It is always \today, today,

\begin{abstract}
From the flashes of fireflies to Josephson junctions and power
infrastructure, networks of coupled phase oscillators provide a powerful framework
to describe synchronization phenomena in many natural and engineered systems. 
Most real-world networks are under the influence of noisy, random inputs, 
potentially inhibiting synchronization. While noise is unavoidable, here we
show that there exist optimal noise patterns which minimize
desynchronizing effects and even enhance order. Specifically, using analytical arguments we show that in the case of a two-oscillator model,
there exists a sharp transition from a regime where the optimal 
synchrony-enhancing noise is perfectly anti-correlated, to one where the optimal
noise is correlated. More generally,
we then use numerical optimization methods to demonstrate that there exist anti-correlated noise patterns
that optimally
enhance synchronization in large complex oscillator networks.
Our results may have implications in networks such as power grids
and neuronal networks,
which are subject to significant amounts of correlated input noise.
\end{abstract}

\maketitle

% \section{Introduction}
The occurrence of noise is unavoidable in networks and systems
at all scales~\cite{Lindner2004}, from biological examples~\cite{Wells2015} such as neurons in the
auditory and visual pathways~\cite{Bialek1987,Tsimring2014,Shlens2008},
neural information processing~\cite{Eyherabide2013,Kanitscheider2015} to
mechanical oscillators~\cite{Matheny2019} and fluctuating inputs affecting the stability of power grids~\cite{Nardelli2014,Milan2013,Schafer2018}.
Synchronization of the underlying network of nonlinear phase oscillators
is a paradigm employed to understand such physical and biological
networks~\cite{Strogatz2012,Mirollo1990}.
Fluctuations are generally seen as undesirable and
significant efforts have been made to understand and prevent their
detrimental
effects on network synchronization~\cite{Luo2007,Bag2007,Yanagita2012,Ronellenfitsch2018a,Hindes2019,Tyloo2018}.
Optimization methods have been successfully employed
to improve synchrony with and without noise, in particular by adjusting the weighted network
topology~\cite{Fazlyab2017,Tanaka2008,Brede2008,Li2017,Kelly2011,Fardad2014,Ronellenfitsch2018a,Skardal2014,Dorfler2012,Alhazmi2019}.
Similar techniques have also been effective for other types
of networks and objectives such as efficient
transport~\cite{Durand2007,Bohn2007,Katifori2010,Rocks2019,Kirkegaard2020,Kaiser2020,Ronellenfitsch2021}.
There has been recent interest in the possibility
that noise may be leveraged to enhance synchronization~\cite{Nicolaou2020,Meng2018,Nagai2010,Nakao2007,Zhou2002,Nakao2016,Aravind2021,Esfahani2012}.
Specifically, it was found that the degree to which input noise
is correlated may have a significant influence on its ability
to aid in or prevent network synchrony~\cite{Nicolaou2020}.

Based on the widely used Kuramoto model~\cite{Kuramoto1984,Acebron2005}, 
here we study the optimal patterns of input noise correlations
that enhance synchronization in networks of oscillators.
Using analytic arguments we find that in the simple case of two
coupled oscillators in the phase-drift regime as studied in
Ref.~\cite{Nicolaou2020}, the optimal synchrony-enhancing noise
undergoes a transition from perfect anti-correlation to
perfect correlation as the total noise strength is increased.
We then numerically study generic complex networks near phase-locked
fixed points and show that the optimal pattern of synchrony-enhancing
noise retains essential characteristics seen
in the two-oscillator case.
The optimal noise we uncover is strongly linked to
the network topology.
In complex networks, the optimal noise correlations show
characteristic clustering, separating the network into regions
that benefit from receiving uncorrelated inputs.
We now proceed to analytically study the tractable case of two
connected Kuramoto oscillators subject to generic noise.

\begin{figure*}
    \centering
    \includegraphics[width=\textwidth]{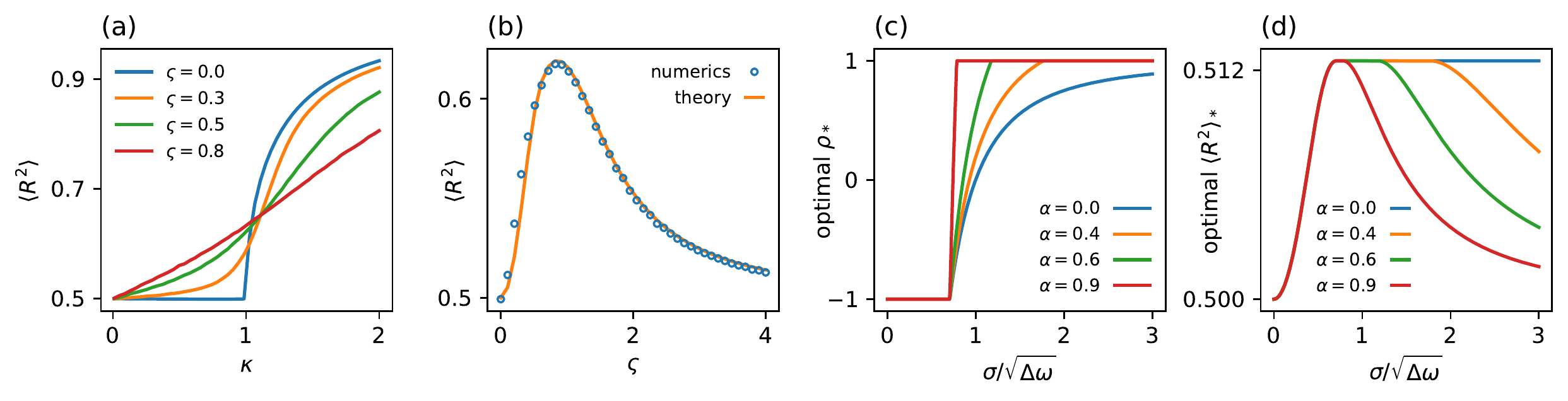}
    \caption{Noise-enhanced synchronization in the two-oscillator model
    and optimal covariance transition.
    (a) Numerically obtained $\langle R^2\rangle$ from integrating Eq.~\eqref{eq:delta-nd} until
    $\tau=400000$ from random initial
    conditions
    as a function of $\kappa=K/\Delta\omega$ for several $\varsigma$.
    (b) Numerically obtained $\langle R^2\rangle$ from integrating Eq.~\eqref{eq:delta-nd}
    until $\tau=20000$ from random
    initial conditions
    (circles) and analytic approximation 
    (see Ref.~\cite{Note1}, Section~I for the explicit formula)
    as a function of $\varsigma$ for $\kappa=0.9$. 
    (c) Numerically approximated optimal covariance $\rho_*$.
    To simplify comparing between different combinations of $\sigma_{1,2}$,
    we introduced the average $\sigma = (\sigma_1 + \sigma_2)/2$
    and the relative difference $\alpha=|\sigma_1 - \sigma_2|/(\sigma_1 + \sigma_2)$.
    Optimal correlations are were obtained at $\kappa = 0.1$. 
    The transitions occur at
    $\sigma_a/\sqrt{\Delta\omega} \approx 1/\sqrt{2}$
    and $\sigma_c/\sqrt{\Delta\omega} \approx 1/(\alpha \sqrt{2})$.
    (d) Approximate optimal order parameter
    $\langle R^2\rangle_*$ for the optimal
    covariances shown in panel (c) at
    $\kappa=0.1$. 
    }
    \label{fig:two-oscillator-synch}
\end{figure*}

% \section{Two-oscillator model}
% \label{sect:two-osc}
The model consists
of coupled phase-oscillators with different natural frequencies.
In the limit of weak coupling, the phases can be modeled using
the Kuramoto-type equations
\begin{align}
    \frac{d\theta_1}{dt} &= \omega_1 + \frac{K}{2}\sin(\theta_2-\theta_1) + \eta_1 \nonumber \\
    \frac{d\theta_2}{dt} &= \omega_2 + \frac{K}{2}\sin(\theta_1-\theta_2) + \eta_2,
    \label{eq:2-osc}
\end{align}
where $\theta_i(t)$ are the oscillator phases, $\omega_i$ the natural
frequencies, $K$ is the coupling constant, and $\eta_i$ are stochastic
white noise terms satisfying $\langle \eta_i\rangle = 0$ and 
$\langle \eta_i(t)
\eta_j(t')\rangle = C_{ij}\, \delta(t-t')$ with the symmetric
and positive semi-definite covariance
matrix $C_{ij} = C_{ji}$.
The model described by Eq.~\eqref{eq:2-osc} was recently shown to
exhibit counter-intuitive enhanced synchronization under 
uncorrelated noise $C_{ij} \sim \delta_{ij}$ as opposed to
common noise $C_{ij} = C$~\cite{Nicolaou2020}.
We now study this effect allowing for arbitrary correlations
between the noise terms.
It is useful to change variables to the mean angle $\mu =
(\theta_1+\theta_2)/2$ and angular difference $\delta =
\theta_1-\theta_2$. The mean $\mu$ is irrelevant for synchronization
as quantified by the squared Kuramoto order parameter $R^2 = |\sum_j e^{i\theta_j}/N|^2 = 1/2 + (1/2)\cos\delta$.
We focus on the equation for the phase difference,
\begin{align}
    \delta'(\tau) = 1 - \kappa \sin\delta(\tau) + \zeta, \label{eq:delta-nd}
\end{align}
where the prime indicates a derivative with respect to $\tau=\Delta\omega\, t$, the dimensionless parameter
$\kappa = K/\Delta\omega$, and the dimensionless
noise $\zeta = (\eta_1 -\eta_2)/\Delta\omega$.

In the absence of noise, $\zeta = 0$, it is well known that
Eq.~\eqref{eq:delta-nd} exhibits a synchronization transition
at $\kappa=1$~\cite{Kuramoto1984,Dorfler2011}. To study non-vanishing noise,
we calculate $\langle \zeta \rangle = 0$ and
\begin{align*}
    \langle \zeta(\tau)\zeta(\tau')\rangle
    &= \frac{1}{\Delta\omega}(C_{11} - 2C_{12} + C_{22})\,
    \delta(\tau-\tau') \\
    &= 2\varsigma^2\, \delta(\tau-\tau').
\end{align*}

This suggests that $\varsigma^2 = (C_{11} - 2C_{12} + C_{22})/(2\Delta\omega)$ is the relevant effective noise strength for synchronization.
This effective noise strength 
depends on the correlation
between the original noise inputs $\eta_{1,2}$. Specifically, 
for common noise, $C_{ij} = \sigma^2$ implies $\varsigma^2 = 0$: Common
noise does not affect synchronization at all. For uncorrelated
noise, $C_{ii} = \sigma^2$ and $C_{12}=0$, which implies
$\varsigma^2 = \sigma^2/\Delta\omega$. A similar argument shows
that the maximum effective noise strength for synchronization
is achieved for \emph{anti-correlated} inputs $C_{ii} = \sigma^2$, $C_{12} = -\sigma^2$ with
$\varsigma^2=2\sigma^2/\Delta\omega$.

But how does $\varsigma^2$ affect
synchronization, and can we find an optimal noise
correlation?
We numerically simulated Eq.~\eqref{eq:delta-nd} and computed
long-time averages of the order parameter $\langle R^2\rangle$ for several
$\varsigma^2$. In the regime below the
transition, $\kappa<1$, noise generally enhances synchronization,
while for $\kappa>1$, noise generally decreases synchronization
\{Fig.~\ref{fig:two-oscillator-synch} (a), Ref.~\cite{Nicolaou2020}\}.
At fixed $\kappa$, there exists an optimal
effective noise $\varsigma_*^2$ that maximizes synchronization
[Fig.~\ref{fig:two-oscillator-synch} (b)]. 

We can relate this to the original
noise covariance matrix as follows.
Fixing the noise variances $C_{ii} = \sigma_i^2$, 
the covariance $C_{12}=\sigma_1 \sigma_2 \rho$ with the
correlation $-1\leq \rho \leq 1$ can be used to
tune the effective noise and thus increase synchronization.
While it appears straightforward to obtain the optimal $\varsigma^2_*$ 
and then to solve $\varsigma^2_* = (\sigma_1^2 -2\sigma_1 \sigma_2 \rho +\sigma_2^2)/(2\Delta\omega)$
for the correlation $\rho$, the constraint $-1\leq \rho \leq 1$
must be taken into account: it is not always possible to adjust $\rho$
and reach the optimal $\varsigma^2_*$. When this happens, the
optimal correlation occurs at the boundary of the allowed range,
$\rho_* = \pm 1$.
\begin{figure}
    \centering
    \includegraphics[width=1.03\columnwidth]{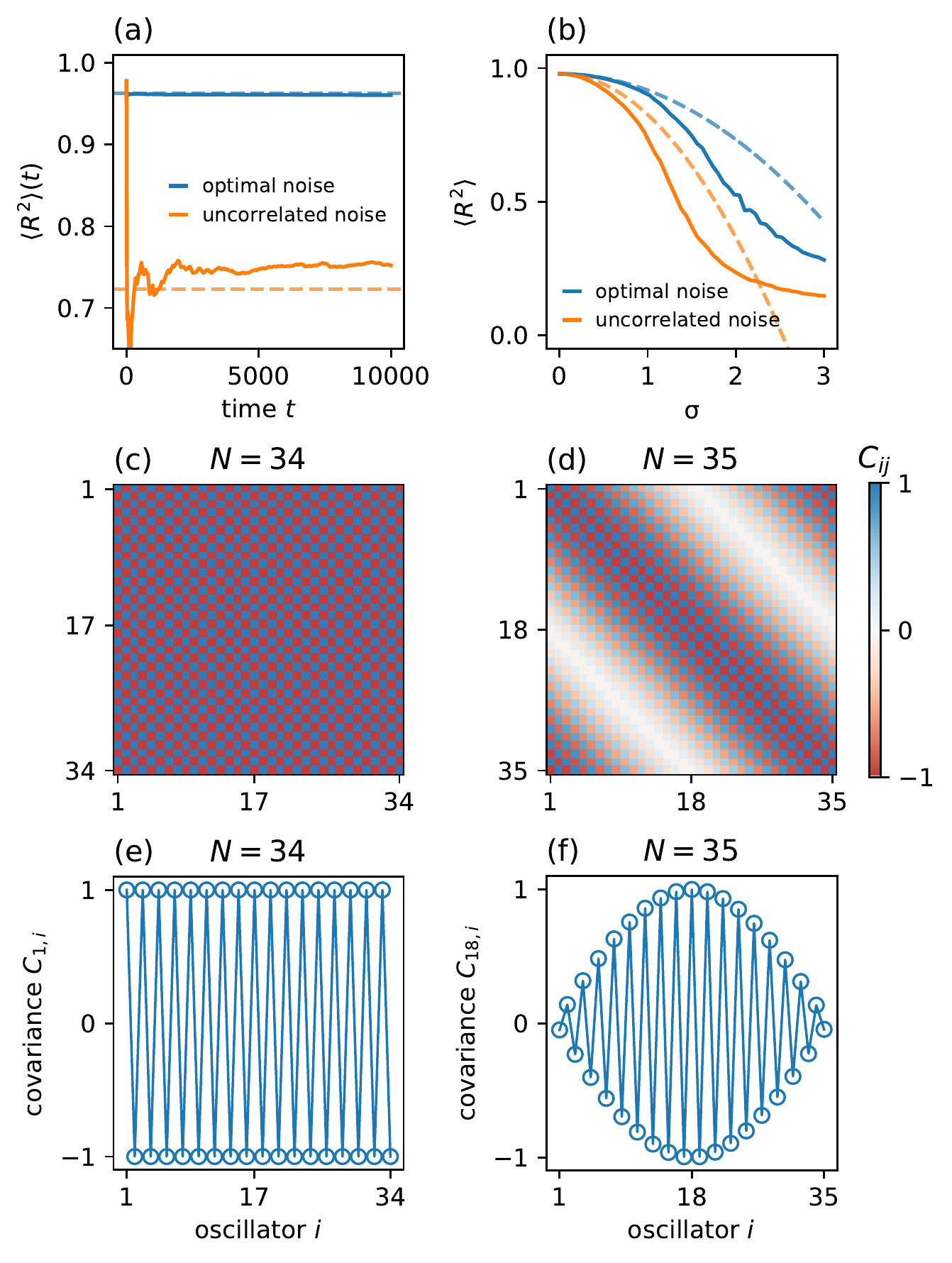}
    \caption{Optimal noise patterns in periodic oscillator chains near fixed points. 
    (a) Time-averaged order parameter
    $\langle{R^2}\rangle(t) = (1/t)\int_0^t R^2(t') dt'$ in a periodic chain of $N=54$ oscillators for optimal and uncorrelated noise. Dashed lines correspond to model predictions from Eq.~\eqref{eq:Rsqr-linear}. The noise variance $\sigma = 0.5$.
    (b) Numerically obtained long-time
    order parameters $\langle{R^2}\rangle$ in
    a $N=20$ periodic chain of oscillators
    with optimal and uncorrelated
    noise. Dashed lines correspond to
    model predictions,
    order parameters were obtained at $t=15000$.
    (c) Optimal covariance matrix for even periodic chain of
    $N=34$.
    (d) Optimal covariance matrix for odd periodic chain of
    $N=35$.
    (e) Optimal covariances $C_{1,i}$ with respect to
    the first oscillator in the even chain.
    (f) Frustrated optimal covariance pattern 
    $C_{18,i}$ with respect
    to the center oscillator in the odd chain.
    Natural frequencies in all panels
    were drawn from the Normal distribution $\mathcal{N}(0,1/N^2)$,
    and $K=2$.
}
    \label{fig:chain-frust}
\end{figure}
Solving for $\rho_*$, the optimal
correlation to enhance synchronization at fixed $\sigma_{1,2}$ 
is then
\begin{align}
    \rho_* = \left[
    \frac{\sigma_1^2 + \sigma_2^2}{2\sigma_1 \sigma_2}
    - \frac{\varsigma_*^2 \Delta\omega}{\sigma_1 \sigma_2}
        \right]',
    \label{eq:rho-opt}
\end{align}
where the primed angle brackets indicate that the argument is clipped to remain
between $-1$ and $1$ using $[x]' = \min(\max(x, -1), 1)$. 
This clipping leads to a sharp transition
[Fig.~\ref{fig:two-oscillator-synch} (c,d)]. 
In the following, it is useful to introduce the average
$\sigma=(\sigma_1 + \sigma_2)/2$.
For small $\sigma/\sqrt{\Delta\omega}$, 
anti-correlated noise optimally enhances synchronization~[Fig.~\ref{fig:two-oscillator-synch} (c)].
Solving Eq.~\eqref{eq:rho-opt} for $\rho_*=-1$, we
find the critical noise strength $\sigma_a$ below which
anti-correlated noise is optimal, $\sigma_a = \varsigma_* \sqrt{\Delta\omega/2}$. Similarly, solving Eq.~\eqref{eq:rho-opt} for $\rho_*=+1$, we find the critical noise strength $\sigma_c$
above which common noise is optimal, $\sigma_c = \sigma_a/\alpha$,
where $\alpha = |\sigma_1 - \sigma_2|/(\sigma_1 + \sigma_2)$.
In the regime where $-1 < \rho_* < 1$, 
the global optimum can be reached and $\langle R^2 \rangle_*$ is constant. Otherwise,
the optimal order parameter occurs at the boundary of
the allowed range of $\rho$ and is less than the global maximum. In this case,
$\rho_* = \pm 1$
[Fig.~\ref{fig:two-oscillator-synch} (c)].

While $\varsigma_*^2$ and thus $\rho_*$ can be obtained numerically,
it is possible to gain insight from an analytic approximation.
Equation~(\ref{eq:delta-nd}) is equivalent to the Fokker-Planck equation
\begin{align}
    \frac{\partial p(\delta,t)}{\partial t} = -\frac{\partial}{\partial \delta}
    \left[ (1 - \kappa \sin\delta)\, p(\delta,t) \right] + \varsigma^2
    \frac{\partial^2p(\delta,t)}{\partial \delta^2} 
    \label{eq:fp}
\end{align}
for the probability density $p(\delta+2\pi,t) = p(\delta,t)$.
From a Fourier series approximation
to the solution of Eq.~\eqref{eq:fp}, we obtain an explicit expression for
$\langle R^2 \rangle(\varsigma, \kappa)$
in the regime of small $\kappa$~\footnote{See Supplemental Material below for the explicit analytical expression for the two-oscillator order parameter, an analysis of the approximations made in the main text, derivations of the two-oscillator model near a fixed point, centered dynamics, a derivation of the Lyapunov equation constraint, explanation of numerical methods, optimal noise for twisted states and periodic square grids, an analysis of the parametric dependence of the optimal noise matrices, and an analysis of second-order power grid models.
The Supplemental Material includes Refs.~\cite{Rackauckas2017,convexjl,Garstka2019,Gajic2008,Wiley2006,Nardelli2014}}.
The optimal effective noise is then
$
    \varsigma_*^2 = 1 -\frac{23}{100} \kappa^2 + \mathcal{O}(\kappa^4),
$
and the corresponding maximal order parameter is
$
    \langle R^2 \rangle_* = \frac{1}{2} +\frac{\kappa }{8}
    + \mathcal{O}(\kappa^3).
$
Even for larger $\kappa \lesssim 1$, there is good agreement with full numerical solutions of the Fokker-Planck equation \{Fig.~\ref{fig:two-oscillator-synch} (b), Ref.~\cite{Note1}, Section~I\}.
The situation is different in the regime $\kappa > 1$ where phase-locked
fixed points exist. 
Here, common noise is always optimal and the noise-free order parameter $R_0^2$ can not be exceeded~\{Ref.~\cite{Note1}, Section~II\}.

% \section{General networks}
In many real-world cases such as
power grids, the network operates near a phase-locked
fixed point instead of in the incoherent regime. Therefore, 
we now focus on general networks near a fixed point and show that, unlike
in the two-oscillator case, there exist optimal correlation patterns
beyond common noise that enhance synchronization. While we note that other
order parameters can also be relevant~\cite{Schroder2017}, here
consider the Kuramoto order parameter.
The equations of motion for $N$ coupled oscillators are
\begin{align}
    \frac{d\theta_i}{dt} = \omega_i + \sum_{j=1}^N K_{ij} \sin(\theta_j-\theta_i)
    + \eta_i, \label{eq:network}
\end{align}
where again the stochastic forcing is given by correlated white noise, 
$\langle \eta_i(t) \eta_j(t')\rangle = C_{ij}\, \delta(t-t')$ and $\langle \eta_i \rangle = 0$.
The matrix $K_{ij} = K_{ji}$ encodes the weighted network topology.
Equation~\eqref{eq:network} as well as the
order parameter allow to shift
the phases as $\theta_i \to \theta_i - \mu$, where $\mu$ is the mean phase. 
In the following, we adopt these ``centered dynamics,'' where
the mean natural frequency, the mean noise, and the mean
covariance with any oscillator vanish, 
$\sum_i \omega_i = \sum_j \eta_j = \sum_j C_{ij} = 0$~\{Ref. \cite{Note1}, Section~III\}.
Note that noise that is uncorrelated 
in the original frame ($C_{ij} \sim \delta_{ij}$) appears uniformly correlated in centered dynamics,
$C_{ij} \sim \delta_{ij} + (\delta_{ij} - 1)/(N-1)$.
Assuming that weak noise drives the centered 
Eq.~\eqref{eq:network} near a fixed point
$
    0 = \omega_i + \sum_{j=1}^N K_{ij} \sin(\bar{\theta}_j-\bar{\theta}_i),
$
we expand $\theta_i = \bar{\theta}_i + \varepsilon_i$
to obtain the linearized dynamics of the perturbations $\varepsilon_i$,
\begin{align}
    \frac{d\varepsilon_i}{dt} = \sum_{j=1}^N K_{ij} \cos(\bar{\theta}_j-\bar{\theta}_i) \left( \varepsilon_j - \varepsilon_i \right) + \eta_i. 
    \label{eq:epsilon}
\end{align}
\begin{figure*}
    \centering
    \includegraphics[width=\textwidth]{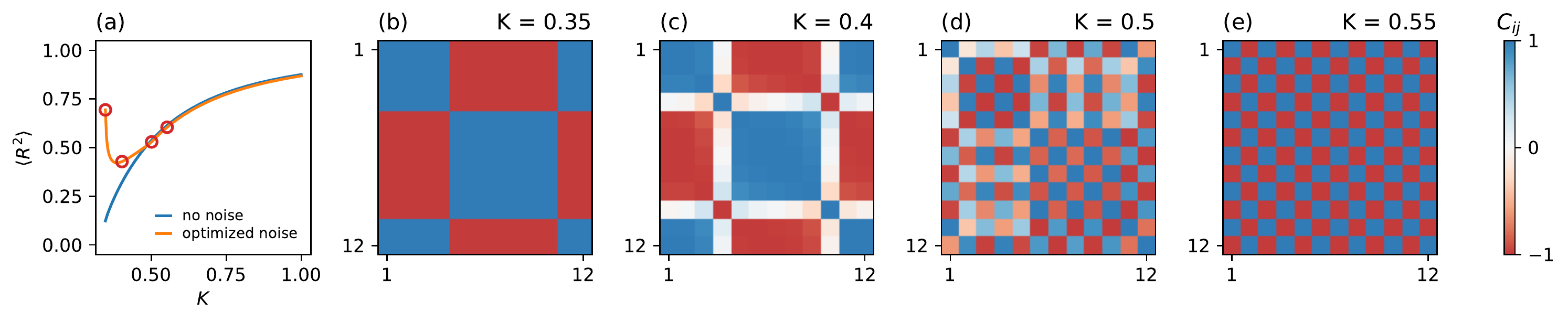}
    \caption{Optimal noise covariance matrix near the 
    synchronization transition in an
    $N=12$ oscillator chain. (a)  Optimal noise
    significantly enhances the approximate order parameter
    $\langle R^2 \rangle = R_0^2 + (1/2) \sigma^2 \operatorname{tr}(HE)$ 
    close to the synchronization transition. Shown are curves
    with $\sigma = 0.25$ and $\sigma = 0$ (no noise).
    (b--e) Optimal covariance matrices near the synchronization transition. The locations corresponding to the matrices
    in panels (b--e) are marked (red circles) in panel (a).
    Close to the phase-drift regime (small $K$), the optimal covariances
    are large-scale ordered and then transition to local order.}
    \label{fig:fig3}
\end{figure*}
In the following, we are interested in long-time averages such that
any initial transients have decayed and the system has settled into an
equilibrium distribution. We expand the order parameter averaged in this way,
\begin{align}
    \langle R^2\rangle = R_0^2 + \frac{1}{2} \langle \bm{\varepsilon}^\top H
    \bm{\varepsilon} \rangle + \mathcal{O}(\varepsilon^3),
    \label{eq:Rsqr-linear}
\end{align}
where $R_0^2$ is the order parameter at the fixed point $\bar{\theta}_i$, and the angle brackets denote
the long-time average.
The linear term $\langle J\bm{\varepsilon} \rangle =  J \langle\bm{\varepsilon}  \rangle$,
where $J$ is the Jacobian matrix, vanishes due to $\langle\bm{\varepsilon} \rangle=0$.
The Hessian matrix
$
    H_{ij} = (2/N^2)\left( \cos(\bar{\theta}_i -\bar{\theta}_j) - \delta_{ij}
    \sum_k \cos(\bar{\theta}_i -\bar{\theta}_k) \right),
$
encodes the synchronization state of the fixed point and is negative
semi-definite close to the synchronous state $\bar{\theta}_i \approx 0$~\{Ref. \cite{Note1}, Section~VI\}, but
can be positive semi-definite or even indefinite, for instance
close to ``twisted states''~\{Refs. \cite{Wiley2006} and \cite{Note1}, Section~VI\}.
Thus, we expect that noise will generically reduce
synchrony. However, it is still possible to find
noise inputs that minimize these effects, and transitions can occur.
Equation~\eqref{eq:Rsqr-linear}
suggests that such optimal synchronization is achieved by maximizing the
second order term 
$
     \langle \bm{\varepsilon}^\top H
    \bm{\varepsilon} \rangle = \operatorname{tr}\left(
    H E\right).
$
Here, $\operatorname{tr}(\cdot)$ is the matrix trace, and $E =
\langle\bm{\varepsilon}\bm{\varepsilon}^\top\rangle$
satisfies the continuous Lyapunov equation $LE + EL = -C$ for the
weighted Laplacian matrix $L_{ij} =  K_{ij} \cos(\bar{\theta}_j-\bar{\theta}_i) - \delta_{ij}\sum_n K_{in} \cos(\bar{\theta}_n-\bar{\theta}_i)$. This can be seen by formally
solving Eq.~\eqref{eq:epsilon} in the Langevin formalism and performing
the noise average~\{Ref. \cite{Note1}, Section~V\}. The Lyapunov equation 
frequently occurs in stability and control theory~\cite{Gajic2008}.
Our goal of finding the optimal noise covariances
$C$ can be formulated as the constrained optimization problem
\begin{align}
    \max_{C,E} &\quad  \operatorname{tr}\left(
    H E\right) \label{eq:opt-prob} \\
\text{such that }&\quad LE + EL = -C \nonumber \\
    &\quad  C \succeq 0. \nonumber
\end{align}
A valid covariance matrix must be positive semi-definite, $C \succeq 0$.
This constraint turns the problem into a semi-definite 
program~\cite{Vandenberghe1996}.
As it stands, the optimization problem is unbounded,
such that we must augment it by an additional constraint to set the noise
scale. 
For simplicity, we fix uniform variances, $C_{ii} = 1$. Because the Lyapunov equation
in Eq.~\eqref{eq:opt-prob} is linear, any uniform $C_{ii}$ can be obtained by rescaling the optimal
solution.
Due to the centered frame constraint $\sum_j C_{ij}=0$ we expect anticorrelations to be relevant
in complex networks again.

% \subsection*{Results}
To uncover the relationship between network topology and
optimal noise, we numerically analyze networks of increasing
complexity.
For simplicity, we take the coupling constants
to be uniform, $K_{ij} \equiv K/d$,
where $d$ is the network's average degree. 
We draw the natural frequencies from a Gaussian distribution
$\omega_i \sim \mathcal{N}(0,1/N^2)$, where $N$ is the number of nodes. 
The mean of the natural frequencies for each network is set to 
exactly zero. Fixed points and solutions to the semi-definite
program Eq.~\eqref{eq:opt-prob} are obtained numerically
\{Ref. \cite{Note1}, Section~IV\}.

As the simplest extension of our two-oscillator model
we first consider periodic chains of $N$ oscillators. 
We note that the optimal noise pattern obtained from
solving Eq.~\eqref{eq:opt-prob} is highly effective in
improving synchronization as compared to uncorrelated noise [Fig.~\ref{fig:chain-frust} (a)],
even far into the nonlinear regime [Fig.~\ref{fig:chain-frust} (b)].
Interestingly, the optimal noise
is such that neighboring pairs in chains with an even
number of oscillators receive anti-correlated
inputs [Fig.~\ref{fig:chain-frust} (c,e)].
However, it is not always possible for
all pairs of neighbors in a network to receive perfectly anti-correlated
inputs, leading to frustrated
patterns of optimal noise.
Indeed, for an odd number of oscillators in the chain,
the magnitude of the optimal noise correlation $|C_{ij}|$ decays 
away from any particular
oscillator $i$. The chain topology prevents any two neighboring oscillators
from receiving perfectly anti-correlated noise [Fig.~\ref{fig:chain-frust} (d,f)].
\begin{figure*}
    \centering
    \includegraphics[width=\textwidth]{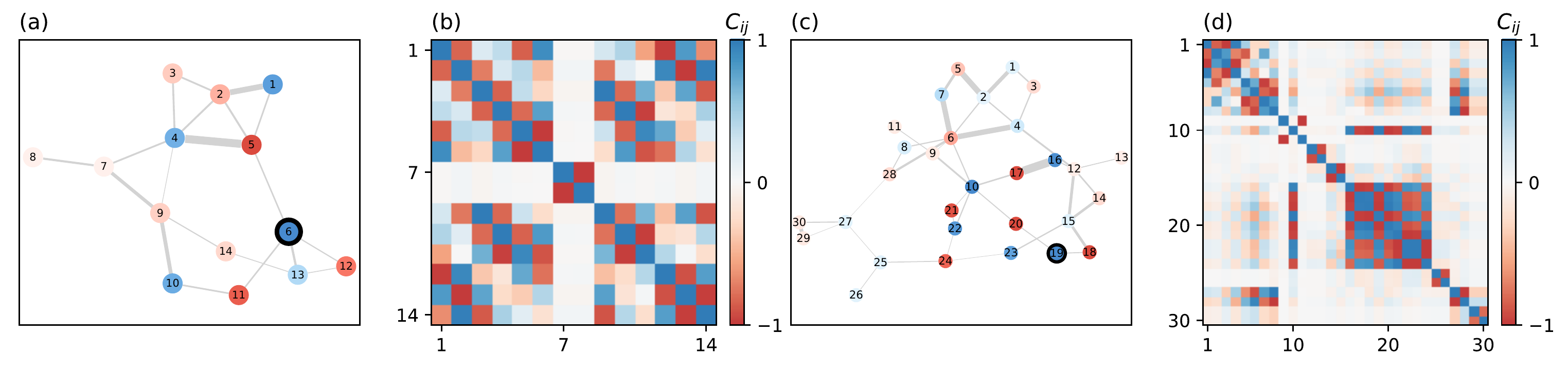}
    \caption{Anti-correlated patterns form clusters 
    in complex networks. (a) IEEE 14-node test grid,
    node colors corresponds to correlation with oscillator
    6.
    (b) Optimal covariance matrix for the network from
    (a). There are two clusters of anti-correlated
    oscillators that are uncorrelated with each other.
    (c) IEEE 30-node test grid,
    node colors corresponds to correlation with oscillator
    19.
    (d) Optimal covariance matrix for the network from
    (c). There are several clusters of anti-correlated
    oscillators that are uncorrelated with each other.
    Small clusters of two nodes generally have one of
    the nodes ``dangling'' with only one neighbor.
    In all graph plots, edge width is proportional to 
    the coupling strength $K_{ij}$.
    Power loads
    were centered and normalized to unit variance, and then
    used as constant inputs.
    }
    \label{fig:ieee}
\end{figure*}
This effect is also seen in periodic grids
\{Ref. \cite{Note1}, Section~VIII\}. It is interesting to note that the optimal noise itself may exhibit a 
transition from local to global organization near the synchronization transition of the underlying network,
depending on the specific set of natural frequencies
$\omega_i$
(Fig.~\ref{fig:fig3}). The transition is accompanied by a significant increase of the order parameter [Fig.~\ref{fig:fig3} (a)]
that even  persists into the phase-drift regime for a large region of couplings $K$~\{Ref. \cite{Note1}, Section~IX\}.

We finally consider complex network topologies derived from power grids~\cite{IEEETest}.
Here, the optimal noise patterns show
clustering, where groups of oscillators are approximately anti-correlated among themselves. 
Correlations between the clusters are approximately zero (Fig.~\ref{fig:ieee}), potentially
promoting cluster synchronization~\cite{Osipov2007}.
One particular type of cluster in these networks consists of one single ``dangling'' node together
with its neighbor. Such nodes have been identified as vulnerable
to perturbations before~\cite{Tyloo2019,Manik2017}. Real power grid dynamics can be modeled
using a second-order model~\cite{Nardelli2014} which is also amenable to our
method and shows a dependence of the optimal covariances on the 
inertia in the network~\{Ref. \cite{Note1}, Section~X\}.

% \section{Discussion}
We studied to what extent synchronization in complex oscillator
networks can be enhanced by correlated noise.
We showed that in the phase-drift regime of a two-oscillator system, 
optimal noise correlations can 
significantly improve
synchronization. The optimal correlations exhibit a transition
between anti-correlated and correlated noise depending on the overall noise strength.
In complex networks, we found that the optimal input noise is generally
anti-correlated with diverse patterns where the strength of correlations
can be constant, decaying, or even be restricted to clusters of oscillators.

Our results may have implications for real networks such as power grids and
neuronal networks. For instance, power grid synchronization may be
enhanced if new power plants and lines are judiciously
placed according to the principles outlined above. Correlations
of input noise can be estimated~\cite{Zhu2021},
or predicted from the weather~\cite{Bett2016,VanderWiel2019}.
Our work opens up new pathways to understanding and
controlling synchronization in complex systems.

\begin{acknowledgments}
S.M., T.S., and T.T.C. acknowledge support from the
Williams College Science Center.
\end{acknowledgments}

\bibliography{noise_synch}% Produces the bibliography via BibTeX.

%Control: production of eprint (1) enabled
\providecommand{\noopsort}[1]{}\providecommand{\singleletter}[1]{#1}%
\begin{thebibliography}{61}%
\makeatletter
\providecommand \@ifxundefined [1]{%
 \@ifx{#1\undefined}
}%
\providecommand \@ifnum [1]{%
 \ifnum #1\expandafter \@firstoftwo
 \else \expandafter \@secondoftwo
 \fi
}%
\providecommand \@ifx [1]{%
 \ifx #1\expandafter \@firstoftwo
 \else \expandafter \@secondoftwo
 \fi
}%
\providecommand \natexlab [1]{#1}%
\providecommand \enquote  [1]{``#1''}%
\providecommand \bibnamefont  [1]{#1}%
\providecommand \bibfnamefont [1]{#1}%
\providecommand \citenamefont [1]{#1}%
\providecommand \href@noop [0]{\@secondoftwo}%
\providecommand \href [0]{\begingroup \@sanitize@url \@href}%
\providecommand \@href[1]{\@@startlink{#1}\@@href}%
\providecommand \@@href[1]{\endgroup#1\@@endlink}%
\providecommand \@sanitize@url [0]{\catcode `\\12\catcode `\$12\catcode
  `\&12\catcode `\#12\catcode `\^12\catcode `\_12\catcode `\%12\relax}%
\providecommand \@@startlink[1]{}%
\providecommand \@@endlink[0]{}%
\providecommand \url  [0]{\begingroup\@sanitize@url \@url }%
\providecommand \@url [1]{\endgroup\@href {#1}{\urlprefix }}%
\providecommand \urlprefix  [0]{URL }%
\providecommand \Eprint [0]{\href }%
\providecommand \doibase [0]{https://doi.org/}%
\providecommand \selectlanguage [0]{\@gobble}%
\providecommand \bibinfo  [0]{\@secondoftwo}%
\providecommand \bibfield  [0]{\@secondoftwo}%
\providecommand \translation [1]{[#1]}%
\providecommand \BibitemOpen [0]{}%
\providecommand \bibitemStop [0]{}%
\providecommand \bibitemNoStop [0]{.\EOS\space}%
\providecommand \EOS [0]{\spacefactor3000\relax}%
\providecommand \BibitemShut  [1]{\csname bibitem#1\endcsname}%
\let\auto@bib@innerbib\@empty
%</preamble>
\bibitem [{\citenamefont {Lindner}(2004)}]{Lindner2004}%
  \BibitemOpen
  \bibfield  {author} {\bibinfo {author} {\bibfnamefont {B.}~\bibnamefont
  {Lindner}},\ }\bibfield  {title} {\bibinfo {title} {{Effects of noise in
  excitable systems}},\ }\href {https://doi.org/10.1016/j.physrep.2003.10.015}
  {\bibfield  {journal} {\bibinfo  {journal} {Physics Reports}\ }\textbf
  {\bibinfo {volume} {392}},\ \bibinfo {pages} {321} (\bibinfo {year}
  {2004})}\BibitemShut {NoStop}%
\bibitem [{\citenamefont {Wells}\ \emph {et~al.}(2015)\citenamefont {Wells},
  \citenamefont {Kath},\ and\ \citenamefont {Motter}}]{Wells2015}%
  \BibitemOpen
  \bibfield  {author} {\bibinfo {author} {\bibfnamefont {D.~K.}\ \bibnamefont
  {Wells}}, \bibinfo {author} {\bibfnamefont {W.~L.}\ \bibnamefont {Kath}},\
  and\ \bibinfo {author} {\bibfnamefont {A.~E.}\ \bibnamefont {Motter}},\
  }\bibfield  {title} {\bibinfo {title} {{Control of Stochastic and Induced
  Switching in Biophysical Networks}},\ }\href
  {https://doi.org/10.1103/PhysRevX.5.031036} {\bibfield  {journal} {\bibinfo
  {journal} {Physical Review X}\ }\textbf {\bibinfo {volume} {5}},\ \bibinfo
  {pages} {031036} (\bibinfo {year} {2015})}\BibitemShut {NoStop}%
\bibitem [{\citenamefont {Bialek}(1987)}]{Bialek1987}%
  \BibitemOpen
  \bibfield  {author} {\bibinfo {author} {\bibfnamefont {W.}~\bibnamefont
  {Bialek}},\ }\bibfield  {title} {\bibinfo {title} {{Physical Limits to
  Sensation and Perception}},\ }\href
  {https://doi.org/10.1146/annurev.bb.16.060187.002323} {\bibfield  {journal}
  {\bibinfo  {journal} {Annual Review of Biophysics and Biophysical Chemistry}\
  }\textbf {\bibinfo {volume} {16}},\ \bibinfo {pages} {455} (\bibinfo {year}
  {1987})}\BibitemShut {NoStop}%
\bibitem [{\citenamefont {Tsimring}(2014)}]{Tsimring2014}%
  \BibitemOpen
  \bibfield  {author} {\bibinfo {author} {\bibfnamefont {L.~S.}\ \bibnamefont
  {Tsimring}},\ }\bibfield  {title} {\bibinfo {title} {{Noise in biology}},\
  }\href {https://doi.org/10.1088/0034-4885/77/2/026601} {\bibfield  {journal}
  {\bibinfo  {journal} {Reports on Progress in Physics}\ }\textbf {\bibinfo
  {volume} {77}},\ \bibinfo {pages} {026601} (\bibinfo {year}
  {2014})}\BibitemShut {NoStop}%
\bibitem [{\citenamefont {Shlens}\ \emph {et~al.}(2008)\citenamefont {Shlens},
  \citenamefont {Rieke},\ and\ \citenamefont {Chichilnisky}}]{Shlens2008}%
  \BibitemOpen
  \bibfield  {author} {\bibinfo {author} {\bibfnamefont {J.}~\bibnamefont
  {Shlens}}, \bibinfo {author} {\bibfnamefont {F.}~\bibnamefont {Rieke}},\ and\
  \bibinfo {author} {\bibfnamefont {E.}~\bibnamefont {Chichilnisky}},\
  }\bibfield  {title} {\bibinfo {title} {{Synchronized firing in the retina}},\
  }\href {https://doi.org/10.1016/j.conb.2008.09.010} {\bibfield  {journal}
  {\bibinfo  {journal} {Current Opinion in Neurobiology}\ }\textbf {\bibinfo
  {volume} {18}},\ \bibinfo {pages} {396} (\bibinfo {year} {2008})}\BibitemShut
  {NoStop}%
\bibitem [{\citenamefont {Eyherabide}\ and\ \citenamefont
  {Samengo}(2013)}]{Eyherabide2013}%
  \BibitemOpen
  \bibfield  {author} {\bibinfo {author} {\bibfnamefont {H.~G.}\ \bibnamefont
  {Eyherabide}}\ and\ \bibinfo {author} {\bibfnamefont {I.}~\bibnamefont
  {Samengo}},\ }\bibfield  {title} {\bibinfo {title} {{When and Why Noise
  Correlations Are Important in Neural Decoding}},\ }\href
  {https://doi.org/10.1523/JNEUROSCI.0357-13.2013} {\bibfield  {journal}
  {\bibinfo  {journal} {Journal of Neuroscience}\ }\textbf {\bibinfo {volume}
  {33}},\ \bibinfo {pages} {17921} (\bibinfo {year} {2013})}\BibitemShut
  {NoStop}%
\bibitem [{\citenamefont {Kanitscheider}\ \emph {et~al.}(2015)\citenamefont
  {Kanitscheider}, \citenamefont {Coen-Cagli},\ and\ \citenamefont
  {Pouget}}]{Kanitscheider2015}%
  \BibitemOpen
  \bibfield  {author} {\bibinfo {author} {\bibfnamefont {I.}~\bibnamefont
  {Kanitscheider}}, \bibinfo {author} {\bibfnamefont {R.}~\bibnamefont
  {Coen-Cagli}},\ and\ \bibinfo {author} {\bibfnamefont {A.}~\bibnamefont
  {Pouget}},\ }\bibfield  {title} {\bibinfo {title} {{Origin of
  information-limiting noise correlations}},\ }\href
  {https://doi.org/10.1073/pnas.1508738112} {\bibfield  {journal} {\bibinfo
  {journal} {Proceedings of the National Academy of Sciences}\ }\textbf
  {\bibinfo {volume} {112}},\ \bibinfo {pages} {E6973} (\bibinfo {year}
  {2015})}\BibitemShut {NoStop}%
\bibitem [{\citenamefont {Matheny}\ \emph {et~al.}(2019)\citenamefont
  {Matheny}, \citenamefont {Emenheiser}, \citenamefont {Fon}, \citenamefont
  {Chapman}, \citenamefont {Salova}, \citenamefont {Rohden}, \citenamefont
  {Li}, \citenamefont {{Hudoba de Badyn}}, \citenamefont {P{\'{o}}sfai},
  \citenamefont {Duenas-Osorio}, \citenamefont {Mesbahi}, \citenamefont
  {Crutchfield}, \citenamefont {Cross}, \citenamefont {D'Souza},\ and\
  \citenamefont {Roukes}}]{Matheny2019}%
  \BibitemOpen
  \bibfield  {author} {\bibinfo {author} {\bibfnamefont {M.~H.}\ \bibnamefont
  {Matheny}}, \bibinfo {author} {\bibfnamefont {J.}~\bibnamefont {Emenheiser}},
  \bibinfo {author} {\bibfnamefont {W.}~\bibnamefont {Fon}}, \bibinfo {author}
  {\bibfnamefont {A.}~\bibnamefont {Chapman}}, \bibinfo {author} {\bibfnamefont
  {A.}~\bibnamefont {Salova}}, \bibinfo {author} {\bibfnamefont
  {M.}~\bibnamefont {Rohden}}, \bibinfo {author} {\bibfnamefont
  {J.}~\bibnamefont {Li}}, \bibinfo {author} {\bibfnamefont {M.}~\bibnamefont
  {{Hudoba de Badyn}}}, \bibinfo {author} {\bibfnamefont {M.}~\bibnamefont
  {P{\'{o}}sfai}}, \bibinfo {author} {\bibfnamefont {L.}~\bibnamefont
  {Duenas-Osorio}}, \bibinfo {author} {\bibfnamefont {M.}~\bibnamefont
  {Mesbahi}}, \bibinfo {author} {\bibfnamefont {J.~P.}\ \bibnamefont
  {Crutchfield}}, \bibinfo {author} {\bibfnamefont {M.~C.}\ \bibnamefont
  {Cross}}, \bibinfo {author} {\bibfnamefont {R.~M.}\ \bibnamefont {D'Souza}},\
  and\ \bibinfo {author} {\bibfnamefont {M.~L.}\ \bibnamefont {Roukes}},\
  }\bibfield  {title} {\bibinfo {title} {{Exotic states in a simple network of
  nanoelectromechanical oscillators}},\ }\href
  {https://doi.org/10.1126/science.aav7932} {\bibfield  {journal} {\bibinfo
  {journal} {Science}\ }\textbf {\bibinfo {volume} {363}},\ \bibinfo {pages}
  {eaav7932} (\bibinfo {year} {2019})}\BibitemShut {NoStop}%
\bibitem [{\citenamefont {Nardelli}\ \emph {et~al.}(2014)\citenamefont
  {Nardelli}, \citenamefont {Rubido}, \citenamefont {Wang}, \citenamefont
  {Baptista}, \citenamefont {Pomalaza-Raez}, \citenamefont {Cardieri},\ and\
  \citenamefont {Latva-aho}}]{Nardelli2014}%
  \BibitemOpen
  \bibfield  {author} {\bibinfo {author} {\bibfnamefont {P.~H.}\ \bibnamefont
  {Nardelli}}, \bibinfo {author} {\bibfnamefont {N.}~\bibnamefont {Rubido}},
  \bibinfo {author} {\bibfnamefont {C.}~\bibnamefont {Wang}}, \bibinfo {author}
  {\bibfnamefont {M.~S.}\ \bibnamefont {Baptista}}, \bibinfo {author}
  {\bibfnamefont {C.}~\bibnamefont {Pomalaza-Raez}}, \bibinfo {author}
  {\bibfnamefont {P.}~\bibnamefont {Cardieri}},\ and\ \bibinfo {author}
  {\bibfnamefont {M.}~\bibnamefont {Latva-aho}},\ }\bibfield  {title} {\bibinfo
  {title} {{Models for the modern power grid}},\ }\href
  {https://doi.org/10.1140/epjst/e2014-02219-6} {\bibfield  {journal} {\bibinfo
   {journal} {The European Physical Journal Special Topics}\ }\textbf {\bibinfo
  {volume} {223}},\ \bibinfo {pages} {2423} (\bibinfo {year}
  {2014})}\BibitemShut {NoStop}%
\bibitem [{\citenamefont {Milan}\ \emph {et~al.}(2013)\citenamefont {Milan},
  \citenamefont {W{\"{a}}chter},\ and\ \citenamefont {Peinke}}]{Milan2013}%
  \BibitemOpen
  \bibfield  {author} {\bibinfo {author} {\bibfnamefont {P.}~\bibnamefont
  {Milan}}, \bibinfo {author} {\bibfnamefont {M.}~\bibnamefont
  {W{\"{a}}chter}},\ and\ \bibinfo {author} {\bibfnamefont {J.}~\bibnamefont
  {Peinke}},\ }\bibfield  {title} {\bibinfo {title} {{Turbulent Character of
  Wind Energy}},\ }\href {https://doi.org/10.1103/PhysRevLett.110.138701}
  {\bibfield  {journal} {\bibinfo  {journal} {Physical Review Letters}\
  }\textbf {\bibinfo {volume} {110}},\ \bibinfo {pages} {138701} (\bibinfo
  {year} {2013})}\BibitemShut {NoStop}%
\bibitem [{\citenamefont {Sch{\"{a}}fer}\ \emph {et~al.}(2018)\citenamefont
  {Sch{\"{a}}fer}, \citenamefont {Beck}, \citenamefont {Aihara}, \citenamefont
  {Witthaut},\ and\ \citenamefont {Timme}}]{Schafer2018}%
  \BibitemOpen
  \bibfield  {author} {\bibinfo {author} {\bibfnamefont {B.}~\bibnamefont
  {Sch{\"{a}}fer}}, \bibinfo {author} {\bibfnamefont {C.}~\bibnamefont {Beck}},
  \bibinfo {author} {\bibfnamefont {K.}~\bibnamefont {Aihara}}, \bibinfo
  {author} {\bibfnamefont {D.}~\bibnamefont {Witthaut}},\ and\ \bibinfo
  {author} {\bibfnamefont {M.}~\bibnamefont {Timme}},\ }\bibfield  {title}
  {\bibinfo {title} {{Non-Gaussian power grid frequency fluctuations
  characterized by L{\'{e}}vy-stable laws and superstatistics}},\ }\href
  {https://doi.org/10.1038/s41560-017-0058-z} {\bibfield  {journal} {\bibinfo
  {journal} {Nature Energy}\ }\textbf {\bibinfo {volume} {3}},\ \bibinfo
  {pages} {119} (\bibinfo {year} {2018})}\BibitemShut {NoStop}%
\bibitem [{\citenamefont {Strogatz}(2012)}]{Strogatz2012}%
  \BibitemOpen
  \bibfield  {author} {\bibinfo {author} {\bibfnamefont {S.}~\bibnamefont
  {Strogatz}},\ }\href {https://books.google.com/books?id=ZQeZAAAAQBAJ} {\emph
  {\bibinfo {title} {Sync: How Order Emerges from Chaos In the Universe,
  Nature, and Daily Life}}}\ (\bibinfo  {publisher} {Hachette Books},\ \bibinfo
  {year} {2012})\BibitemShut {NoStop}%
\bibitem [{\citenamefont {Mirollo}\ and\ \citenamefont
  {Strogatz}(1990)}]{Mirollo1990}%
  \BibitemOpen
  \bibfield  {author} {\bibinfo {author} {\bibfnamefont {R.~E.}\ \bibnamefont
  {Mirollo}}\ and\ \bibinfo {author} {\bibfnamefont {S.~H.}\ \bibnamefont
  {Strogatz}},\ }\bibfield  {title} {\bibinfo {title} {{Synchronization of
  Pulse-Coupled Biological Oscillators}},\ }\href
  {https://doi.org/10.1137/0150098} {\bibfield  {journal} {\bibinfo  {journal}
  {SIAM Journal on Applied Mathematics}\ }\textbf {\bibinfo {volume} {50}},\
  \bibinfo {pages} {1645} (\bibinfo {year} {1990})}\BibitemShut {NoStop}%
\bibitem [{\citenamefont {Luo}\ and\ \citenamefont {Banakar}(2007)}]{Luo2007}%
  \BibitemOpen
  \bibfield  {author} {\bibinfo {author} {\bibfnamefont {C.}~\bibnamefont
  {Luo}}\ and\ \bibinfo {author} {\bibfnamefont {H.}~\bibnamefont {Banakar}},\
  }\bibfield  {title} {\bibinfo {title} {{Strategies to smooth wind power
  fluctuations of wind turbine generator}},\ }\href
  {http://ieeexplore.ieee.org/xpls/abs_all.jsp?arnumber=4207451} {\bibfield
  {journal} {\bibinfo  {journal} {IEEE Transactions on Energy Conversion}\
  }\textbf {\bibinfo {volume} {22}},\ \bibinfo {pages} {341} (\bibinfo {year}
  {2007})}\BibitemShut {NoStop}%
\bibitem [{\citenamefont {Bag}\ \emph {et~al.}(2007)\citenamefont {Bag},
  \citenamefont {Petrosyan},\ and\ \citenamefont {Hu}}]{Bag2007}%
  \BibitemOpen
  \bibfield  {author} {\bibinfo {author} {\bibfnamefont {B.~C.}\ \bibnamefont
  {Bag}}, \bibinfo {author} {\bibfnamefont {K.~G.}\ \bibnamefont {Petrosyan}},\
  and\ \bibinfo {author} {\bibfnamefont {C.-K.}\ \bibnamefont {Hu}},\
  }\bibfield  {title} {\bibinfo {title} {{Influence of noise on the
  synchronization of the stochastic Kuramoto model}},\ }\href
  {https://doi.org/10.1103/PhysRevE.76.056210} {\bibfield  {journal} {\bibinfo
  {journal} {Physical Review E}\ }\textbf {\bibinfo {volume} {76}},\ \bibinfo
  {pages} {056210} (\bibinfo {year} {2007})}\BibitemShut {NoStop}%
\bibitem [{\citenamefont {Yanagita}\ and\ \citenamefont
  {Mikhailov}(2012)}]{Yanagita2012}%
  \BibitemOpen
  \bibfield  {author} {\bibinfo {author} {\bibfnamefont {T.}~\bibnamefont
  {Yanagita}}\ and\ \bibinfo {author} {\bibfnamefont {A.~S.}\ \bibnamefont
  {Mikhailov}},\ }\bibfield  {title} {\bibinfo {title} {{Design of oscillator
  networks with enhanced synchronization tolerance against noise}},\ }\href
  {https://doi.org/10.1103/PhysRevE.85.056206} {\bibfield  {journal} {\bibinfo
  {journal} {Physical Review E}\ }\textbf {\bibinfo {volume} {85}},\ \bibinfo
  {pages} {056206} (\bibinfo {year} {2012})}\BibitemShut {NoStop}%
\bibitem [{\citenamefont {Ronellenfitsch}\ \emph {et~al.}(2018)\citenamefont
  {Ronellenfitsch}, \citenamefont {Dunkel},\ and\ \citenamefont
  {Wilczek}}]{Ronellenfitsch2018a}%
  \BibitemOpen
  \bibfield  {author} {\bibinfo {author} {\bibfnamefont {H.}~\bibnamefont
  {Ronellenfitsch}}, \bibinfo {author} {\bibfnamefont {J.}~\bibnamefont
  {Dunkel}},\ and\ \bibinfo {author} {\bibfnamefont {M.}~\bibnamefont
  {Wilczek}},\ }\bibfield  {title} {\bibinfo {title} {{Optimal Noise-Canceling
  Networks}},\ }\href {https://doi.org/10.1103/PhysRevLett.121.208301}
  {\bibfield  {journal} {\bibinfo  {journal} {Physical Review Letters}\
  }\textbf {\bibinfo {volume} {121}},\ \bibinfo {pages} {208301} (\bibinfo
  {year} {2018})}\BibitemShut {NoStop}%
\bibitem [{\citenamefont {Hindes}\ \emph {et~al.}(2019)\citenamefont {Hindes},
  \citenamefont {Jacquod},\ and\ \citenamefont {Schwartz}}]{Hindes2019}%
  \BibitemOpen
  \bibfield  {author} {\bibinfo {author} {\bibfnamefont {J.}~\bibnamefont
  {Hindes}}, \bibinfo {author} {\bibfnamefont {P.}~\bibnamefont {Jacquod}},\
  and\ \bibinfo {author} {\bibfnamefont {I.~B.}\ \bibnamefont {Schwartz}},\
  }\bibfield  {title} {\bibinfo {title} {{Network desynchronization by
  non-Gaussian fluctuations}},\ }\href
  {https://doi.org/10.1103/PhysRevE.100.052314} {\bibfield  {journal} {\bibinfo
   {journal} {Physical Review E}\ }\textbf {\bibinfo {volume} {100}},\ \bibinfo
  {pages} {052314} (\bibinfo {year} {2019})}\BibitemShut {NoStop}%
\bibitem [{\citenamefont {Tyloo}\ \emph {et~al.}(2018)\citenamefont {Tyloo},
  \citenamefont {Coletta},\ and\ \citenamefont {Jacquod}}]{Tyloo2018}%
  \BibitemOpen
  \bibfield  {author} {\bibinfo {author} {\bibfnamefont {M.}~\bibnamefont
  {Tyloo}}, \bibinfo {author} {\bibfnamefont {T.}~\bibnamefont {Coletta}},\
  and\ \bibinfo {author} {\bibfnamefont {P.}~\bibnamefont {Jacquod}},\
  }\bibfield  {title} {\bibinfo {title} {{Robustness of Synchrony in Complex
  Networks and Generalized Kirchhoff Indices}},\ }\href
  {https://doi.org/10.1103/PhysRevLett.120.084101} {\bibfield  {journal}
  {\bibinfo  {journal} {Physical Review Letters}\ }\textbf {\bibinfo {volume}
  {120}},\ \bibinfo {pages} {084101} (\bibinfo {year} {2018})}\BibitemShut
  {NoStop}%
\bibitem [{\citenamefont {Fazlyab}\ \emph {et~al.}(2017)\citenamefont
  {Fazlyab}, \citenamefont {D{\"{o}}rfler},\ and\ \citenamefont
  {Preciado}}]{Fazlyab2017}%
  \BibitemOpen
  \bibfield  {author} {\bibinfo {author} {\bibfnamefont {M.}~\bibnamefont
  {Fazlyab}}, \bibinfo {author} {\bibfnamefont {F.}~\bibnamefont
  {D{\"{o}}rfler}},\ and\ \bibinfo {author} {\bibfnamefont {V.~M.}\
  \bibnamefont {Preciado}},\ }\bibfield  {title} {\bibinfo {title} {{Optimal
  network design for synchronization of coupled oscillators}},\ }\href
  {https://doi.org/10.1016/j.automatica.2017.07.005} {\bibfield  {journal}
  {\bibinfo  {journal} {Automatica}\ }\textbf {\bibinfo {volume} {84}},\
  \bibinfo {pages} {181} (\bibinfo {year} {2017})}\BibitemShut {NoStop}%
\bibitem [{\citenamefont {Tanaka}\ and\ \citenamefont
  {Aoyagi}(2008)}]{Tanaka2008}%
  \BibitemOpen
  \bibfield  {author} {\bibinfo {author} {\bibfnamefont {T.}~\bibnamefont
  {Tanaka}}\ and\ \bibinfo {author} {\bibfnamefont {T.}~\bibnamefont
  {Aoyagi}},\ }\bibfield  {title} {\bibinfo {title} {{Optimal weighted networks
  of phase oscillators for synchronization}},\ }\href
  {https://doi.org/10.1103/PhysRevE.78.046210} {\bibfield  {journal} {\bibinfo
  {journal} {Physical Review E}\ }\textbf {\bibinfo {volume} {78}},\ \bibinfo
  {pages} {046210} (\bibinfo {year} {2008})}\BibitemShut {NoStop}%
\bibitem [{\citenamefont {Brede}(2008)}]{Brede2008}%
  \BibitemOpen
  \bibfield  {author} {\bibinfo {author} {\bibfnamefont {M.}~\bibnamefont
  {Brede}},\ }\bibfield  {title} {\bibinfo {title} {{Synchrony-optimized
  networks of non-identical Kuramoto oscillators}},\ }\href
  {https://doi.org/10.1016/j.physleta.2007.11.069} {\bibfield  {journal}
  {\bibinfo  {journal} {Physics Letters A}\ }\textbf {\bibinfo {volume}
  {372}},\ \bibinfo {pages} {2618} (\bibinfo {year} {2008})}\BibitemShut
  {NoStop}%
\bibitem [{\citenamefont {Li}\ and\ \citenamefont {Wong}(2017)}]{Li2017}%
  \BibitemOpen
  \bibfield  {author} {\bibinfo {author} {\bibfnamefont {B.}~\bibnamefont
  {Li}}\ and\ \bibinfo {author} {\bibfnamefont {K.~Y.~M.}\ \bibnamefont
  {Wong}},\ }\bibfield  {title} {\bibinfo {title} {{Optimizing synchronization
  stability of the Kuramoto model in complex networks and power grids}},\
  }\href {https://doi.org/10.1103/PhysRevE.95.012207} {\bibfield  {journal}
  {\bibinfo  {journal} {Physical Review E}\ }\textbf {\bibinfo {volume} {95}},\
  \bibinfo {pages} {012207} (\bibinfo {year} {2017})}\BibitemShut {NoStop}%
\bibitem [{\citenamefont {Kelly}\ and\ \citenamefont
  {Gottwald}(2011)}]{Kelly2011}%
  \BibitemOpen
  \bibfield  {author} {\bibinfo {author} {\bibfnamefont {D.}~\bibnamefont
  {Kelly}}\ and\ \bibinfo {author} {\bibfnamefont {G.~A.}\ \bibnamefont
  {Gottwald}},\ }\bibfield  {title} {\bibinfo {title} {{On the topology of
  synchrony optimized networks of a Kuramoto-model with non-identical
  oscillators}},\ }\href {https://doi.org/10.1063/1.3590855} {\bibfield
  {journal} {\bibinfo  {journal} {Chaos: An Interdisciplinary Journal of
  Nonlinear Science}\ }\textbf {\bibinfo {volume} {21}},\ \bibinfo {pages}
  {025110} (\bibinfo {year} {2011})}\BibitemShut {NoStop}%
\bibitem [{\citenamefont {Fardad}\ \emph {et~al.}(2014)\citenamefont {Fardad},
  \citenamefont {Lin},\ and\ \citenamefont {Jovanovic}}]{Fardad2014}%
  \BibitemOpen
  \bibfield  {author} {\bibinfo {author} {\bibfnamefont {M.}~\bibnamefont
  {Fardad}}, \bibinfo {author} {\bibfnamefont {F.}~\bibnamefont {Lin}},\ and\
  \bibinfo {author} {\bibfnamefont {M.~R.}\ \bibnamefont {Jovanovic}},\
  }\bibfield  {title} {\bibinfo {title} {{Design of Optimal Sparse
  Interconnection Graphs for Synchronization of Oscillator Networks}},\ }\href
  {https://doi.org/10.1109/TAC.2014.2301577} {\bibfield  {journal} {\bibinfo
  {journal} {IEEE Transactions on Automatic Control}\ }\textbf {\bibinfo
  {volume} {59}},\ \bibinfo {pages} {2457} (\bibinfo {year}
  {2014})}\BibitemShut {NoStop}%
\bibitem [{\citenamefont {Skardal}\ \emph {et~al.}(2014)\citenamefont
  {Skardal}, \citenamefont {Taylor},\ and\ \citenamefont {Sun}}]{Skardal2014}%
  \BibitemOpen
  \bibfield  {author} {\bibinfo {author} {\bibfnamefont {P.~S.}\ \bibnamefont
  {Skardal}}, \bibinfo {author} {\bibfnamefont {D.}~\bibnamefont {Taylor}},\
  and\ \bibinfo {author} {\bibfnamefont {J.}~\bibnamefont {Sun}},\ }\bibfield
  {title} {\bibinfo {title} {{Optimal synchronization of complex networks}},\
  }\href {https://doi.org/10.1103/PhysRevLett.113.144101} {\bibfield  {journal}
  {\bibinfo  {journal} {Physical Review Letters}\ }\textbf {\bibinfo {volume}
  {113}},\ \bibinfo {pages} {1} (\bibinfo {year} {2014})}\BibitemShut {NoStop}%
\bibitem [{\citenamefont {D{\"{o}}rfler}\ and\ \citenamefont
  {Bullo}(2012)}]{Dorfler2012}%
  \BibitemOpen
  \bibfield  {author} {\bibinfo {author} {\bibfnamefont {F.}~\bibnamefont
  {D{\"{o}}rfler}}\ and\ \bibinfo {author} {\bibfnamefont {F.}~\bibnamefont
  {Bullo}},\ }\bibfield  {title} {\bibinfo {title} {{Synchronization and
  Transient Stability in Power Networks and Nonuniform Kuramoto Oscillators}},\
  }\href {https://doi.org/10.1137/110851584} {\bibfield  {journal} {\bibinfo
  {journal} {SIAM Journal on Control and Optimization}\ }\textbf {\bibinfo
  {volume} {50}},\ \bibinfo {pages} {1616} (\bibinfo {year}
  {2012})}\BibitemShut {NoStop}%
\bibitem [{\citenamefont {Alhazmi}\ \emph {et~al.}(2019)\citenamefont
  {Alhazmi}, \citenamefont {Dehghanian}, \citenamefont {Wang},\ and\
  \citenamefont {Shinde}}]{Alhazmi2019}%
  \BibitemOpen
  \bibfield  {author} {\bibinfo {author} {\bibfnamefont {M.}~\bibnamefont
  {Alhazmi}}, \bibinfo {author} {\bibfnamefont {P.}~\bibnamefont {Dehghanian}},
  \bibinfo {author} {\bibfnamefont {S.}~\bibnamefont {Wang}},\ and\ \bibinfo
  {author} {\bibfnamefont {B.}~\bibnamefont {Shinde}},\ }\bibfield  {title}
  {\bibinfo {title} {{Power Grid Optimal Topology Control Considering
  Correlations of System Uncertainties}},\ }\href
  {https://doi.org/10.1109/TIA.2019.2934706} {\bibfield  {journal} {\bibinfo
  {journal} {IEEE Transactions on Industry Applications}\ }\textbf {\bibinfo
  {volume} {55}},\ \bibinfo {pages} {5594} (\bibinfo {year}
  {2019})}\BibitemShut {NoStop}%
\bibitem [{\citenamefont {Durand}(2007)}]{Durand2007}%
  \BibitemOpen
  \bibfield  {author} {\bibinfo {author} {\bibfnamefont {M.}~\bibnamefont
  {Durand}},\ }\bibfield  {title} {\bibinfo {title} {{Structure of Optimal
  Transport Networks Subject to a Global Constraint}},\ }\href
  {https://doi.org/10.1103/PhysRevLett.98.088701} {\bibfield  {journal}
  {\bibinfo  {journal} {Physical Review Letters}\ }\textbf {\bibinfo {volume}
  {98}},\ \bibinfo {pages} {088701} (\bibinfo {year} {2007})}\BibitemShut
  {NoStop}%
\bibitem [{\citenamefont {Bohn}\ and\ \citenamefont
  {Magnasco}(2007)}]{Bohn2007}%
  \BibitemOpen
  \bibfield  {author} {\bibinfo {author} {\bibfnamefont {S.}~\bibnamefont
  {Bohn}}\ and\ \bibinfo {author} {\bibfnamefont {M.~O.}\ \bibnamefont
  {Magnasco}},\ }\bibfield  {title} {\bibinfo {title} {{Structure, Scaling, and
  Phase Transition in the Optimal Transport Network}},\ }\href
  {https://doi.org/10.1103/PhysRevLett.98.088702} {\bibfield  {journal}
  {\bibinfo  {journal} {Physical Review Letters}\ }\textbf {\bibinfo {volume}
  {98}},\ \bibinfo {pages} {088702} (\bibinfo {year} {2007})}\BibitemShut
  {NoStop}%
\bibitem [{\citenamefont {Katifori}\ \emph {et~al.}(2010)\citenamefont
  {Katifori}, \citenamefont {Sz{\"{o}}llősi},\ and\ \citenamefont
  {Magnasco}}]{Katifori2010}%
  \BibitemOpen
  \bibfield  {author} {\bibinfo {author} {\bibfnamefont {E.}~\bibnamefont
  {Katifori}}, \bibinfo {author} {\bibfnamefont {G.~J.}\ \bibnamefont
  {Sz{\"{o}}llősi}},\ and\ \bibinfo {author} {\bibfnamefont {M.~O.}\
  \bibnamefont {Magnasco}},\ }\bibfield  {title} {\bibinfo {title} {{Damage and
  Fluctuations Induce Loops in Optimal Transport Networks}},\ }\href
  {https://doi.org/10.1103/PhysRevLett.104.048704} {\bibfield  {journal}
  {\bibinfo  {journal} {Physical Review Letters}\ }\textbf {\bibinfo {volume}
  {104}},\ \bibinfo {pages} {048704} (\bibinfo {year} {2010})}\BibitemShut
  {NoStop}%
\bibitem [{\citenamefont {Rocks}\ \emph {et~al.}(2019)\citenamefont {Rocks},
  \citenamefont {Ronellenfitsch}, \citenamefont {Liu}, \citenamefont {Nagel},\
  and\ \citenamefont {Katifori}}]{Rocks2019}%
  \BibitemOpen
  \bibfield  {author} {\bibinfo {author} {\bibfnamefont {J.~W.}\ \bibnamefont
  {Rocks}}, \bibinfo {author} {\bibfnamefont {H.}~\bibnamefont
  {Ronellenfitsch}}, \bibinfo {author} {\bibfnamefont {A.~J.}\ \bibnamefont
  {Liu}}, \bibinfo {author} {\bibfnamefont {S.~R.}\ \bibnamefont {Nagel}},\
  and\ \bibinfo {author} {\bibfnamefont {E.}~\bibnamefont {Katifori}},\
  }\bibfield  {title} {\bibinfo {title} {{Limits of multifunctionality in
  tunable networks}},\ }\href {https://doi.org/10.1073/pnas.1806790116}
  {\bibfield  {journal} {\bibinfo  {journal} {Proceedings of the National
  Academy of Sciences}\ }\textbf {\bibinfo {volume} {116}},\ \bibinfo {pages}
  {2506} (\bibinfo {year} {2019})}\BibitemShut {NoStop}%
\bibitem [{\citenamefont {Kirkegaard}\ and\ \citenamefont
  {Sneppen}(2020)}]{Kirkegaard2020}%
  \BibitemOpen
  \bibfield  {author} {\bibinfo {author} {\bibfnamefont {J.~B.}\ \bibnamefont
  {Kirkegaard}}\ and\ \bibinfo {author} {\bibfnamefont {K.}~\bibnamefont
  {Sneppen}},\ }\bibfield  {title} {\bibinfo {title} {{Optimal Transport Flows
  for Distributed Production Networks}},\ }\href
  {https://doi.org/10.1103/PhysRevLett.124.208101} {\bibfield  {journal}
  {\bibinfo  {journal} {Physical Review Letters}\ }\textbf {\bibinfo {volume}
  {124}},\ \bibinfo {pages} {208101} (\bibinfo {year} {2020})}\BibitemShut
  {NoStop}%
\bibitem [{\citenamefont {Kaiser}\ \emph {et~al.}(2020)\citenamefont {Kaiser},
  \citenamefont {Ronellenfitsch},\ and\ \citenamefont {Witthaut}}]{Kaiser2020}%
  \BibitemOpen
  \bibfield  {author} {\bibinfo {author} {\bibfnamefont {F.}~\bibnamefont
  {Kaiser}}, \bibinfo {author} {\bibfnamefont {H.}~\bibnamefont
  {Ronellenfitsch}},\ and\ \bibinfo {author} {\bibfnamefont {D.}~\bibnamefont
  {Witthaut}},\ }\bibfield  {title} {\bibinfo {title} {{Discontinuous
  transition to loop formation in optimal supply networks}},\ }\href
  {https://doi.org/10.1038/s41467-020-19567-2} {\bibfield  {journal} {\bibinfo
  {journal} {Nature Communications}\ }\textbf {\bibinfo {volume} {11}},\
  \bibinfo {pages} {5796} (\bibinfo {year} {2020})}\BibitemShut {NoStop}%
\bibitem [{\citenamefont {Ronellenfitsch}(2021)}]{Ronellenfitsch2021}%
  \BibitemOpen
  \bibfield  {author} {\bibinfo {author} {\bibfnamefont {H.}~\bibnamefont
  {Ronellenfitsch}},\ }\bibfield  {title} {\bibinfo {title} {{Optimal
  Elasticity of Biological Networks}},\ }\href
  {https://doi.org/10.1103/PhysRevLett.126.038101} {\bibfield  {journal}
  {\bibinfo  {journal} {Physical Review Letters}\ }\textbf {\bibinfo {volume}
  {126}},\ \bibinfo {pages} {038101} (\bibinfo {year} {2021})}\BibitemShut
  {NoStop}%
\bibitem [{\citenamefont {Nicolaou}\ \emph {et~al.}(2020)\citenamefont
  {Nicolaou}, \citenamefont {Sebek}, \citenamefont {Kiss},\ and\ \citenamefont
  {Motter}}]{Nicolaou2020}%
  \BibitemOpen
  \bibfield  {author} {\bibinfo {author} {\bibfnamefont {Z.~G.}\ \bibnamefont
  {Nicolaou}}, \bibinfo {author} {\bibfnamefont {M.}~\bibnamefont {Sebek}},
  \bibinfo {author} {\bibfnamefont {I.~Z.}\ \bibnamefont {Kiss}},\ and\
  \bibinfo {author} {\bibfnamefont {A.~E.}\ \bibnamefont {Motter}},\ }\bibfield
   {title} {\bibinfo {title} {{Coherent Dynamics Enhanced by Uncorrelated
  Noise}},\ }\href {https://doi.org/10.1103/PhysRevLett.125.094101} {\bibfield
  {journal} {\bibinfo  {journal} {Physical Review Letters}\ }\textbf {\bibinfo
  {volume} {125}},\ \bibinfo {pages} {094101} (\bibinfo {year}
  {2020})}\BibitemShut {NoStop}%
\bibitem [{\citenamefont {Meng}\ and\ \citenamefont {Riecke}(2018)}]{Meng2018}%
  \BibitemOpen
  \bibfield  {author} {\bibinfo {author} {\bibfnamefont {J.~H.}\ \bibnamefont
  {Meng}}\ and\ \bibinfo {author} {\bibfnamefont {H.}~\bibnamefont {Riecke}},\
  }\bibfield  {title} {\bibinfo {title} {{Synchronization by uncorrelated
  noise: interacting rhythms in interconnected oscillator networks}},\ }\href
  {https://doi.org/10.1038/s41598-018-24670-y} {\bibfield  {journal} {\bibinfo
  {journal} {Scientific Reports}\ }\textbf {\bibinfo {volume} {8}},\ \bibinfo
  {pages} {6949} (\bibinfo {year} {2018})}\BibitemShut {NoStop}%
\bibitem [{\citenamefont {Nagai}\ and\ \citenamefont {Kori}(2010)}]{Nagai2010}%
  \BibitemOpen
  \bibfield  {author} {\bibinfo {author} {\bibfnamefont {K.~H.}\ \bibnamefont
  {Nagai}}\ and\ \bibinfo {author} {\bibfnamefont {H.}~\bibnamefont {Kori}},\
  }\bibfield  {title} {\bibinfo {title} {{Noise-induced synchronization of a
  large population of globally coupled nonidentical oscillators}},\ }\href
  {https://doi.org/10.1103/PhysRevE.81.065202} {\bibfield  {journal} {\bibinfo
  {journal} {Physical Review E}\ }\textbf {\bibinfo {volume} {81}},\ \bibinfo
  {pages} {065202} (\bibinfo {year} {2010})}\BibitemShut {NoStop}%
\bibitem [{\citenamefont {Nakao}\ \emph {et~al.}(2007)\citenamefont {Nakao},
  \citenamefont {Arai},\ and\ \citenamefont {Kawamura}}]{Nakao2007}%
  \BibitemOpen
  \bibfield  {author} {\bibinfo {author} {\bibfnamefont {H.}~\bibnamefont
  {Nakao}}, \bibinfo {author} {\bibfnamefont {K.}~\bibnamefont {Arai}},\ and\
  \bibinfo {author} {\bibfnamefont {Y.}~\bibnamefont {Kawamura}},\ }\bibfield
  {title} {\bibinfo {title} {{Noise-Induced Synchronization and Clustering in
  Ensembles of Uncoupled Limit-Cycle Oscillators}},\ }\href
  {https://doi.org/10.1103/PhysRevLett.98.184101} {\bibfield  {journal}
  {\bibinfo  {journal} {Physical Review Letters}\ }\textbf {\bibinfo {volume}
  {98}},\ \bibinfo {pages} {184101} (\bibinfo {year} {2007})}\BibitemShut
  {NoStop}%
\bibitem [{\citenamefont {Zhou}\ \emph {et~al.}(2002)\citenamefont {Zhou},
  \citenamefont {Kurths}, \citenamefont {Kiss},\ and\ \citenamefont
  {Hudson}}]{Zhou2002}%
  \BibitemOpen
  \bibfield  {author} {\bibinfo {author} {\bibfnamefont {C.}~\bibnamefont
  {Zhou}}, \bibinfo {author} {\bibfnamefont {J.}~\bibnamefont {Kurths}},
  \bibinfo {author} {\bibfnamefont {I.~Z.}\ \bibnamefont {Kiss}},\ and\
  \bibinfo {author} {\bibfnamefont {J.~L.}\ \bibnamefont {Hudson}},\ }\bibfield
   {title} {\bibinfo {title} {{Noise-Enhanced Phase Synchronization of Chaotic
  Oscillators}},\ }\href {https://doi.org/10.1103/PhysRevLett.89.014101}
  {\bibfield  {journal} {\bibinfo  {journal} {Physical Review Letters}\
  }\textbf {\bibinfo {volume} {89}},\ \bibinfo {pages} {014101} (\bibinfo
  {year} {2002})}\BibitemShut {NoStop}%
\bibitem [{\citenamefont {Nakao}(2016)}]{Nakao2016}%
  \BibitemOpen
  \bibfield  {author} {\bibinfo {author} {\bibfnamefont {H.}~\bibnamefont
  {Nakao}},\ }\bibfield  {title} {\bibinfo {title} {{Phase reduction approach
  to synchronisation of nonlinear oscillators}},\ }\href
  {https://doi.org/10.1080/00107514.2015.1094987} {\bibfield  {journal}
  {\bibinfo  {journal} {Contemporary Physics}\ }\textbf {\bibinfo {volume}
  {57}},\ \bibinfo {pages} {188} (\bibinfo {year} {2016})}\BibitemShut
  {NoStop}%
\bibitem [{\citenamefont {Aravind}\ \emph {et~al.}(2021)\citenamefont
  {Aravind}, \citenamefont {Sinha},\ and\ \citenamefont
  {Parmananda}}]{Aravind2021}%
  \BibitemOpen
  \bibfield  {author} {\bibinfo {author} {\bibfnamefont {M.}~\bibnamefont
  {Aravind}}, \bibinfo {author} {\bibfnamefont {S.}~\bibnamefont {Sinha}},\
  and\ \bibinfo {author} {\bibfnamefont {P.}~\bibnamefont {Parmananda}},\
  }\bibfield  {title} {\bibinfo {title} {{Competitive interplay of repulsive
  coupling and cross-correlated noises in bistable systems}},\ }\href
  {https://doi.org/10.1063/5.0056173} {\bibfield  {journal} {\bibinfo
  {journal} {Chaos: An Interdisciplinary Journal of Nonlinear Science}\
  }\textbf {\bibinfo {volume} {31}},\ \bibinfo {pages} {061106} (\bibinfo
  {year} {2021})}\BibitemShut {NoStop}%
\bibitem [{\citenamefont {Esfahani}\ \emph {et~al.}(2012)\citenamefont
  {Esfahani}, \citenamefont {Shahbazi},\ and\ \citenamefont
  {Samani}}]{Esfahani2012}%
  \BibitemOpen
  \bibfield  {author} {\bibinfo {author} {\bibfnamefont {R.~K.}\ \bibnamefont
  {Esfahani}}, \bibinfo {author} {\bibfnamefont {F.}~\bibnamefont {Shahbazi}},\
  and\ \bibinfo {author} {\bibfnamefont {K.~A.}\ \bibnamefont {Samani}},\
  }\bibfield  {title} {\bibinfo {title} {{Noise-induced synchronization in
  small world networks of phase oscillators}},\ }\href
  {https://doi.org/10.1103/PhysRevE.86.036204} {\bibfield  {journal} {\bibinfo
  {journal} {Physical Review E}\ }\textbf {\bibinfo {volume} {86}},\ \bibinfo
  {pages} {036204} (\bibinfo {year} {2012})}\BibitemShut {NoStop}%
\bibitem [{\citenamefont {Kuramoto}(1984)}]{Kuramoto1984}%
  \BibitemOpen
  \bibfield  {author} {\bibinfo {author} {\bibfnamefont {Y.}~\bibnamefont
  {Kuramoto}},\ }\href {https://doi.org/10.1007/978-3-642-69689-3} {\emph
  {\bibinfo {title} {{Chemical Oscillations, Waves, and Turbulence}}}},\
  \bibinfo {series} {Springer Series in Synergetics}, Vol.~\bibinfo {volume}
  {19}\ (\bibinfo  {publisher} {Springer},\ \bibinfo {address} {Berlin,
  Heidelberg},\ \bibinfo {year} {1984})\BibitemShut {NoStop}%
\bibitem [{\citenamefont {Acebr{\'{o}}n}\ \emph {et~al.}(2005)\citenamefont
  {Acebr{\'{o}}n}, \citenamefont {Bonilla}, \citenamefont {{P{\'{e}}rez
  Vicente}}, \citenamefont {Ritort},\ and\ \citenamefont
  {Spigler}}]{Acebron2005}%
  \BibitemOpen
  \bibfield  {author} {\bibinfo {author} {\bibfnamefont {J.~A.}\ \bibnamefont
  {Acebr{\'{o}}n}}, \bibinfo {author} {\bibfnamefont {L.~L.}\ \bibnamefont
  {Bonilla}}, \bibinfo {author} {\bibfnamefont {C.~J.}\ \bibnamefont
  {{P{\'{e}}rez Vicente}}}, \bibinfo {author} {\bibfnamefont {F.}~\bibnamefont
  {Ritort}},\ and\ \bibinfo {author} {\bibfnamefont {R.}~\bibnamefont
  {Spigler}},\ }\bibfield  {title} {\bibinfo {title} {{The Kuramoto model: A
  simple paradigm for synchronization phenomena}},\ }\href
  {https://doi.org/10.1103/RevModPhys.77.137} {\bibfield  {journal} {\bibinfo
  {journal} {Reviews of Modern Physics}\ }\textbf {\bibinfo {volume} {77}},\
  \bibinfo {pages} {137} (\bibinfo {year} {2005})}\BibitemShut {NoStop}%
\bibitem [{Note1()}]{Note1}%
  \BibitemOpen
  \bibinfo {note} {See Supplemental Material below for the explicit analytical
  expression for the two-oscillator order parameter, an analysis of the
  approximations made in the main text, derivations of the two-oscillator model
  near a fixed point, centered dynamics, a derivation of the Lyapunov equation
  constraint, explanation of numerical methods, optimal noise for twisted
  states and periodic square grids, an analysis of the parametric dependence of
  the optimal noise matrices, and an analysis of second-order power grid
  models. The Supplemental Material includes Refs.~\cite
  {Rackauckas2017,convexjl,Garstka2019,Gajic2008,Wiley2006,Nardelli2014}}\BibitemShut
  {NoStop}%
\bibitem [{\citenamefont {D{\"{o}}rfler}\ and\ \citenamefont
  {Bullo}(2011)}]{Dorfler2011}%
  \BibitemOpen
  \bibfield  {author} {\bibinfo {author} {\bibfnamefont {F.}~\bibnamefont
  {D{\"{o}}rfler}}\ and\ \bibinfo {author} {\bibfnamefont {F.}~\bibnamefont
  {Bullo}},\ }\bibfield  {title} {\bibinfo {title} {{On the Critical Coupling
  for Kuramoto Oscillators}},\ }\href {https://doi.org/10.1137/10081530X}
  {\bibfield  {journal} {\bibinfo  {journal} {SIAM Journal on Applied Dynamical
  Systems}\ }\textbf {\bibinfo {volume} {10}},\ \bibinfo {pages} {1070}
  (\bibinfo {year} {2011})}\BibitemShut {NoStop}%
\bibitem [{\citenamefont {Schr{\"{o}}der}\ \emph {et~al.}(2017)\citenamefont
  {Schr{\"{o}}der}, \citenamefont {Timme},\ and\ \citenamefont
  {Witthaut}}]{Schroder2017}%
  \BibitemOpen
  \bibfield  {author} {\bibinfo {author} {\bibfnamefont {M.}~\bibnamefont
  {Schr{\"{o}}der}}, \bibinfo {author} {\bibfnamefont {M.}~\bibnamefont
  {Timme}},\ and\ \bibinfo {author} {\bibfnamefont {D.}~\bibnamefont
  {Witthaut}},\ }\bibfield  {title} {\bibinfo {title} {{A universal order
  parameter for synchrony in networks of limit cycle oscillators}},\ }\href
  {https://doi.org/10.1063/1.4995963} {\bibfield  {journal} {\bibinfo
  {journal} {Chaos: An Interdisciplinary Journal of Nonlinear Science}\
  }\textbf {\bibinfo {volume} {27}},\ \bibinfo {pages} {073119} (\bibinfo
  {year} {2017})}\BibitemShut {NoStop}%
\bibitem [{\citenamefont {Wiley}\ \emph {et~al.}(2006)\citenamefont {Wiley},
  \citenamefont {Strogatz},\ and\ \citenamefont {Girvan}}]{Wiley2006}%
  \BibitemOpen
  \bibfield  {author} {\bibinfo {author} {\bibfnamefont {D.~A.}\ \bibnamefont
  {Wiley}}, \bibinfo {author} {\bibfnamefont {S.~H.}\ \bibnamefont
  {Strogatz}},\ and\ \bibinfo {author} {\bibfnamefont {M.}~\bibnamefont
  {Girvan}},\ }\bibfield  {title} {\bibinfo {title} {{The size of the sync
  basin}},\ }\href {https://doi.org/10.1063/1.2165594} {\bibfield  {journal}
  {\bibinfo  {journal} {Chaos: An Interdisciplinary Journal of Nonlinear
  Science}\ }\textbf {\bibinfo {volume} {16}},\ \bibinfo {pages} {015103}
  (\bibinfo {year} {2006})}\BibitemShut {NoStop}%
\bibitem [{\citenamefont {Gaji{\'c}}\ and\ \citenamefont
  {Qureshi}(2008)}]{Gajic2008}%
  \BibitemOpen
  \bibfield  {author} {\bibinfo {author} {\bibfnamefont {Z.}~\bibnamefont
  {Gaji{\'c}}}\ and\ \bibinfo {author} {\bibfnamefont {M.~T.~J.}\ \bibnamefont
  {Qureshi}},\ }\href@noop {} {\emph {\bibinfo {title} {{Lyapunov Matrix
  Equation in System Stability and Control}}}}\ (\bibinfo  {publisher} {Dover
  Publications},\ \bibinfo {address} {Mineola, NY},\ \bibinfo {year} {2008})\
  p.\ \bibinfo {pages} {272}\BibitemShut {NoStop}%
\bibitem [{\citenamefont {Vandenberghe}\ and\ \citenamefont
  {Boyd}(1996)}]{Vandenberghe1996}%
  \BibitemOpen
  \bibfield  {author} {\bibinfo {author} {\bibfnamefont {L.}~\bibnamefont
  {Vandenberghe}}\ and\ \bibinfo {author} {\bibfnamefont {S.}~\bibnamefont
  {Boyd}},\ }\bibfield  {title} {\bibinfo {title} {{Semidefinite
  Programming}},\ }\href {https://doi.org/10.1137/1038003} {\bibfield
  {journal} {\bibinfo  {journal} {SIAM Review}\ }\textbf {\bibinfo {volume}
  {38}},\ \bibinfo {pages} {49} (\bibinfo {year} {1996})}\BibitemShut {NoStop}%
\bibitem [{\citenamefont {{University of Washington}}()}]{IEEETest}%
  \BibitemOpen
  \bibfield  {author} {\bibinfo {author} {\bibnamefont {{University of
  Washington}}},\ }\href {https://labs.ece.uw.edu/pstca/} {\bibinfo {title}
  {Power systems test case archive}}\BibitemShut {NoStop}%
\bibitem [{\citenamefont {Osipov}\ \emph {et~al.}(2007)\citenamefont {Osipov},
  \citenamefont {Kurths},\ and\ \citenamefont {Zhou}}]{Osipov2007}%
  \BibitemOpen
  \bibfield  {author} {\bibinfo {author} {\bibfnamefont {G.~V.}\ \bibnamefont
  {Osipov}}, \bibinfo {author} {\bibfnamefont {J.}~\bibnamefont {Kurths}},\
  and\ \bibinfo {author} {\bibfnamefont {C.}~\bibnamefont {Zhou}},\ }\href@noop
  {} {\emph {\bibinfo {title} {{Synchronization in Oscillatory Networks}}}}\
  (\bibinfo  {publisher} {Springer-Verlag},\ \bibinfo {address} {Berlin
  Heidelberg},\ \bibinfo {year} {2007})\ p.\ \bibinfo {pages} {373}\BibitemShut
  {NoStop}%
\bibitem [{\citenamefont {Tyloo}\ \emph {et~al.}(2019)\citenamefont {Tyloo},
  \citenamefont {Pagnier},\ and\ \citenamefont {Jacquod}}]{Tyloo2019}%
  \BibitemOpen
  \bibfield  {author} {\bibinfo {author} {\bibfnamefont {M.}~\bibnamefont
  {Tyloo}}, \bibinfo {author} {\bibfnamefont {L.}~\bibnamefont {Pagnier}},\
  and\ \bibinfo {author} {\bibfnamefont {P.}~\bibnamefont {Jacquod}},\
  }\bibfield  {title} {\bibinfo {title} {{The key player problem in complex
  oscillator networks and electric power grids: Resistance centralities
  identify local vulnerabilities}},\ }\href
  {https://doi.org/10.1126/sciadv.aaw8359} {\bibfield  {journal} {\bibinfo
  {journal} {Science Advances}\ }\textbf {\bibinfo {volume} {5}},\ \bibinfo
  {pages} {eaaw8359} (\bibinfo {year} {2019})}\BibitemShut {NoStop}%
\bibitem [{\citenamefont {Manik}\ \emph {et~al.}(2017)\citenamefont {Manik},
  \citenamefont {Rohden}, \citenamefont {Ronellenfitsch}, \citenamefont
  {Zhang}, \citenamefont {Hallerberg}, \citenamefont {Witthaut},\ and\
  \citenamefont {Timme}}]{Manik2017}%
  \BibitemOpen
  \bibfield  {author} {\bibinfo {author} {\bibfnamefont {D.}~\bibnamefont
  {Manik}}, \bibinfo {author} {\bibfnamefont {M.}~\bibnamefont {Rohden}},
  \bibinfo {author} {\bibfnamefont {H.}~\bibnamefont {Ronellenfitsch}},
  \bibinfo {author} {\bibfnamefont {X.}~\bibnamefont {Zhang}}, \bibinfo
  {author} {\bibfnamefont {S.}~\bibnamefont {Hallerberg}}, \bibinfo {author}
  {\bibfnamefont {D.}~\bibnamefont {Witthaut}},\ and\ \bibinfo {author}
  {\bibfnamefont {M.}~\bibnamefont {Timme}},\ }\bibfield  {title} {\bibinfo
  {title} {{Network susceptibilities: Theory and applications}},\ }\href
  {https://doi.org/10.1103/PhysRevE.95.012319} {\bibfield  {journal} {\bibinfo
  {journal} {Physical Review E}\ }\textbf {\bibinfo {volume} {95}},\ \bibinfo
  {pages} {012319} (\bibinfo {year} {2017})}\BibitemShut {NoStop}%
\bibitem [{\citenamefont {Zhu}\ and\ \citenamefont {Lin}(2021)}]{Zhu2021}%
  \BibitemOpen
  \bibfield  {author} {\bibinfo {author} {\bibfnamefont {L.}~\bibnamefont
  {Zhu}}\ and\ \bibinfo {author} {\bibfnamefont {J.}~\bibnamefont {Lin}},\
  }\bibfield  {title} {\bibinfo {title} {{Learning Spatiotemporal Correlations
  for Missing Noisy PMU Data Correction in Smart Grid}},\ }\href
  {https://doi.org/10.1109/JIOT.2020.3040195} {\bibfield  {journal} {\bibinfo
  {journal} {IEEE Internet of Things Journal}\ }\textbf {\bibinfo {volume}
  {8}},\ \bibinfo {pages} {7589} (\bibinfo {year} {2021})}\BibitemShut
  {NoStop}%
\bibitem [{\citenamefont {Bett}\ and\ \citenamefont
  {Thornton}(2016)}]{Bett2016}%
  \BibitemOpen
  \bibfield  {author} {\bibinfo {author} {\bibfnamefont {P.~E.}\ \bibnamefont
  {Bett}}\ and\ \bibinfo {author} {\bibfnamefont {H.~E.}\ \bibnamefont
  {Thornton}},\ }\bibfield  {title} {\bibinfo {title} {{The climatological
  relationships between wind and solar energy supply in Britain}},\ }\href
  {https://doi.org/10.1016/j.renene.2015.10.006} {\bibfield  {journal}
  {\bibinfo  {journal} {Renewable Energy}\ }\textbf {\bibinfo {volume} {87}},\
  \bibinfo {pages} {96} (\bibinfo {year} {2016})}\BibitemShut {NoStop}%
\bibitem [{\citenamefont {van~der Wiel}\ \emph {et~al.}(2019)\citenamefont
  {van~der Wiel}, \citenamefont {Bloomfield}, \citenamefont {Lee},
  \citenamefont {Stoop}, \citenamefont {Blackport}, \citenamefont {Screen},\
  and\ \citenamefont {Selten}}]{VanderWiel2019}%
  \BibitemOpen
  \bibfield  {author} {\bibinfo {author} {\bibfnamefont {K.}~\bibnamefont
  {van~der Wiel}}, \bibinfo {author} {\bibfnamefont {H.~C.}\ \bibnamefont
  {Bloomfield}}, \bibinfo {author} {\bibfnamefont {R.~W.}\ \bibnamefont {Lee}},
  \bibinfo {author} {\bibfnamefont {L.~P.}\ \bibnamefont {Stoop}}, \bibinfo
  {author} {\bibfnamefont {R.}~\bibnamefont {Blackport}}, \bibinfo {author}
  {\bibfnamefont {J.~A.}\ \bibnamefont {Screen}},\ and\ \bibinfo {author}
  {\bibfnamefont {F.~M.}\ \bibnamefont {Selten}},\ }\bibfield  {title}
  {\bibinfo {title} {{The influence of weather regimes on European renewable
  energy production and demand}},\ }\href
  {https://doi.org/10.1088/1748-9326/ab38d3} {\bibfield  {journal} {\bibinfo
  {journal} {Environmental Research Letters}\ }\textbf {\bibinfo {volume}
  {14}},\ \bibinfo {pages} {094010} (\bibinfo {year} {2019})}\BibitemShut
  {NoStop}%
\bibitem [{\citenamefont {Rackauckas}\ and\ \citenamefont
  {Nie}(2017)}]{Rackauckas2017}%
  \BibitemOpen
  \bibfield  {author} {\bibinfo {author} {\bibfnamefont {C.}~\bibnamefont
  {Rackauckas}}\ and\ \bibinfo {author} {\bibfnamefont {Q.}~\bibnamefont
  {Nie}},\ }\bibfield  {title} {\bibinfo {title} {{DifferentialEquations.jl}--a
  performant and feature-rich ecosystem for solving differential equations in
  {Julia}},\ }\href {http://doi.org/10.5334/jors.151} {\bibfield  {journal}
  {\bibinfo  {journal} {Journal of Open Research Software}\ }\textbf {\bibinfo
  {volume} {5}} (\bibinfo {year} {2017})}\BibitemShut {NoStop}%
\bibitem [{\citenamefont {Udell}\ \emph {et~al.}(2014)\citenamefont {Udell},
  \citenamefont {Mohan}, \citenamefont {Zeng}, \citenamefont {Hong},
  \citenamefont {Diamond},\ and\ \citenamefont {Boyd}}]{convexjl}%
  \BibitemOpen
  \bibfield  {author} {\bibinfo {author} {\bibfnamefont {M.}~\bibnamefont
  {Udell}}, \bibinfo {author} {\bibfnamefont {K.}~\bibnamefont {Mohan}},
  \bibinfo {author} {\bibfnamefont {D.}~\bibnamefont {Zeng}}, \bibinfo {author}
  {\bibfnamefont {J.}~\bibnamefont {Hong}}, \bibinfo {author} {\bibfnamefont
  {S.}~\bibnamefont {Diamond}},\ and\ \bibinfo {author} {\bibfnamefont
  {S.}~\bibnamefont {Boyd}},\ }\bibfield  {title} {\bibinfo {title} {Convex
  optimization in {Julia}},\ }in\ \href {https://doi.org/10.1109/HPTCDL.2014.5}
  {\emph {\bibinfo {booktitle} {2014 First Workshop for High Performance
  Technical Computing in Dynamic Languages}}}\ (\bibinfo {year} {2014})\ pp.\
  \bibinfo {pages} {18--28}\BibitemShut {NoStop}%
\bibitem [{\citenamefont {Garstka}\ \emph {et~al.}(2021)\citenamefont
  {Garstka}, \citenamefont {Cannon},\ and\ \citenamefont
  {Goulart}}]{Garstka2019}%
  \BibitemOpen
  \bibfield  {author} {\bibinfo {author} {\bibfnamefont {M.}~\bibnamefont
  {Garstka}}, \bibinfo {author} {\bibfnamefont {M.}~\bibnamefont {Cannon}},\
  and\ \bibinfo {author} {\bibfnamefont {P.}~\bibnamefont {Goulart}},\
  }\bibfield  {title} {\bibinfo {title} {{COSMO: A Conic Operator Splitting
  Method for Convex Conic Problems}},\ }\href
  {https://doi.org/10.1007/s10957-021-01896-x} {\bibfield  {journal} {\bibinfo
  {journal} {Journal of Optimization Theory and Applications}\ }\textbf
  {\bibinfo {volume} {190}},\ \bibinfo {pages} {779} (\bibinfo {year}
  {2021})}\BibitemShut {NoStop}%
\end{thebibliography}%

\onecolumngrid

\beginsupplement

\newpage

\begin{center}
    \textbf{SUPPLEMENTAL MATERIAL}
\end{center}

% Supplement goes here

\section{Approximate solution of the Fokker-Planck equation}
While Eq.~\eqref{eq:fp} can not be solved analytically, in the regime of $\kappa\ll 1$, the
steady-state distribution appears to be well-represented by sinusoids. 
Encouraged by this, we approximate the steady-state distribution
as a truncated Fourier series, $p(\delta) \approx \sum_{k=-N}^N a_k\,
e^{ik\delta}$. The normalization condition $\int_{-\pi}^\pi
p(\delta)\,d\delta=1$ implies $a_0=1/(2\pi)$ and reality of the solution
is equivalent to $a_{-k} = a_k^*$.
Plugging in the shortest non-trivial Fourier series with $N=2$ and
solving the resulting system of equations yields
the following approximation for the Fourier coefficients,
\begin{align*}
    a_0 &= \frac{1}{2\pi} \\
    a_1 &= \frac{1}{2\pi} \frac{2 \kappa  \left(2\varsigma ^2+i\right)}{\kappa ^2+8 \varsigma ^4+12 i \varsigma ^2-4} \\
    a_2 &= \frac{1}{2\pi} \frac{\kappa ^2}{\kappa ^2+8 \varsigma ^4+12 i \varsigma ^2-4}.
\end{align*}
Using this Fourier approximation, we can calculate the expected order
parameter as
\begin{align}
    \langle R^2\rangle &= \frac{1}{2} + \frac{1}{2} \langle\cos\delta\rangle 
    = \frac{1}{2} + \frac{\pi}{2}(a_1 + a_1^*) \nonumber\\
    &= \frac{1}{2} + \frac{2 \kappa  \varsigma ^2 \left(\kappa ^2+8 \varsigma ^4+2\right)}{\left(\kappa ^2-4\right)^2+16 \left(\kappa ^2+5\right) \varsigma ^4+64 \varsigma ^8}
\end{align}
where the expectation value is defined as $\langle f(\delta)\rangle = \int_{-\pi}^\pi f(\delta) p(\delta)\, d\delta$.
This equation can be optimized directly with respect to $\varsigma$ by setting the derivative to zero and solving in Mathematica.

\begin{figure}[h]
    \centering
    \includegraphics[width=0.45\textwidth]{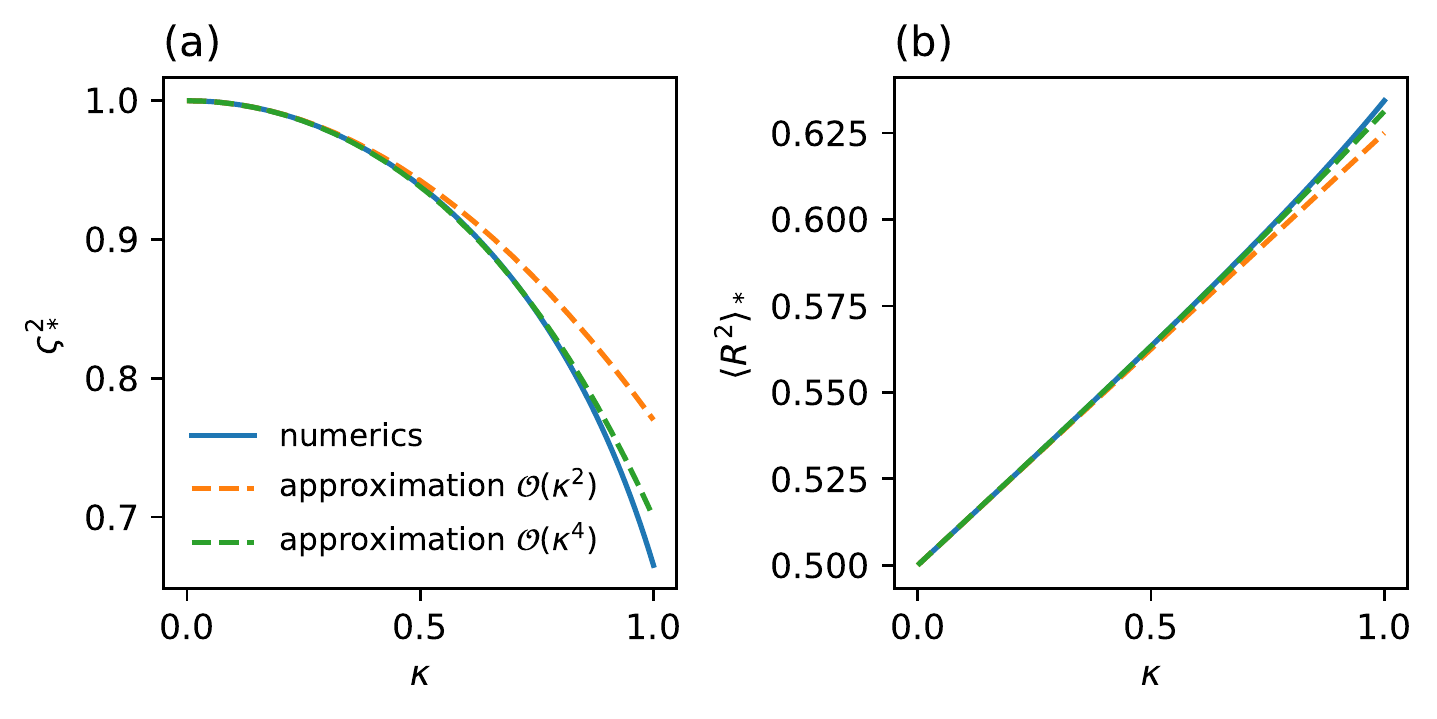}
    \caption{Comparison of the approximation formulas derived in the main text to
    corresponding numerically obtained optimal values from a full numerical solution
    of the Fokker-Planck equation Eq.~\eqref{eq:fp}.
    (a) Optimal effective noise strength where the approximation formula is
    $\varsigma^2_* \approx 1 - 23\kappa^2/100 - 1757\kappa^4/25000$.
    Note that in the main paper, only the expression to order
    $\mathcal{O}(\kappa^2)$ is given.
    (b) Optimal order parameter where the approximation formula is
    $\langle R^2\rangle_* \approx 1/2 + \kappa/8 + \kappa^3/160$.
    Note that in the main paper, only the expression to order
    $\mathcal{O}(\kappa)$ is given. The colored lines have the
    same interpretation as in panel (a).}
    \label{fig:approximation}
\end{figure}

The first approximation found above can be compared to numerical solutions.
Specifically, we are interested in the optimal effective noise strength $\varsigma_*^2$
and the corresponding optimal order parameter $\langle R^2 \rangle_*$.
We solved the Fokker-Planck equation Eq.~\eqref{eq:fp} numerically using Mathematica
and numerically obtained $\varsigma_*^2$ and $\langle R^2 \rangle_*$ using a
Golden Section search (termination accuracy was set to $10^{-6}$) as a function of $\kappa$.
Except for values of $\kappa$ close to 1, the approximation formulas are very good
(Fig.~\ref{fig:approximation}).

% It should be noted that it is possible to obtain other, inequivalent approximate solutions
% to the Fokker-Planck equation as well. For instance, one can use the integral representation
% (see Ref.~\cite{Nicolaou2020}),
% \begin{align*}
%     p(\phi) = A \int_0^{2\pi} d\psi \left[
%     \frac{1}{\exp(2\pi /\varsigma^2) - 1} + H(\phi-\psi) 
%     \right] e^{ (\psi - \phi)/\varsigma^2}
%     e^{- (\cos\psi - \cos\phi)\kappa/\varsigma^2},
% \end{align*}
% where $A$ is a normalization factor ensuring $\int_0^{2\pi} p(\phi) = 1$,
% $H(x)$ is the Heaviside step function, and $\phi=-\delta$.
% It is natural to expand the second exponential,
% \begin{align*}
%     e^{- (\cos\psi - \cos\phi)\kappa/\varsigma^2} = 1 - \frac{\kappa}{\varsigma^2} (\cos\psi - \cos\phi) + \frac{1}{2} \frac{\kappa^2}{\varsigma^4} (\cos\psi - \cos\phi)^2 - \dots
% \end{align*}
% Each term in the resulting series can be integrated in closed form, and an approximation
% to $p(\phi)$ is thus obtained. However, the approximation is in $\kappa/\varsigma^2$ and is
% therefore, unlike the Fourier approximation, not valid in the limit of $\varsigma \to 0$.

\section{Two-oscillator model near a fixed point}
Here, we analyze the two-oscillator model near a fixed point. We
start by computing the fixed point in the noise-free case,
\begin{align*}
    \delta' = 0 \Rightarrow \bar\delta = \arcsin ( \kappa^{-1} ).
\end{align*}
We now expand the equation of motion with noise close to the fixed point, 
$\delta(\tau) = \bar\delta + \varepsilon(\tau)$ as
\begin{align}
    \delta' = \varepsilon' &= 1 - \kappa \sin(\bar\delta + \varepsilon) + \zeta \nonumber \\
    &\approx -\sqrt{1-\kappa^{-2}}\, \varepsilon + \zeta. \label{eq:two-eps}
\end{align}
We can similarly expand the order parameter,
\begin{align}
    R^2 &= \frac{1}{2} + \frac{1}{2} \cos(\bar\delta + \varepsilon) \nonumber \\
    &= R_0^2 - \frac{1}{2\kappa} \varepsilon - \frac{1}{4} \sqrt{1-\kappa^{-2}} \, \varepsilon^2 + \mathcal{O}(\varepsilon^3), \label{eq:two-rsqr}
\end{align}
where $R_0^2 = (1/2)(1 + \sqrt{1-\kappa^{-2}})$. Averaging Eq.~\eqref{eq:two-rsqr}
over the noise, the term linear in $\varepsilon$ drops out and we are left
with
\begin{align*}
    \langle R^2 \rangle = R_0^2 - \frac{1}{4} \sqrt{1-\kappa^{-2}} \, \langle \varepsilon^2
    \rangle.
\end{align*}
To calculate the average $\langle \varepsilon^2\rangle$, we formally
solve Eq.~\eqref{eq:two-eps} as
\begin{align*}
    \varepsilon(\tau) = \varepsilon_0\,  e^{-\sqrt{1-\kappa^{-2}}\tau} + e^{-\sqrt{1-\kappa^{-2}}\tau}
    \int_0^\tau ds \, e^{\sqrt{1-\kappa^{-2}}s} \zeta(s).
\end{align*}
For long times, the first term vanishes and we can use the second term to calculate
\begin{align*}
    \langle \varepsilon(\tau)^2 \rangle 
    &= e^{-2 \sqrt{1 - \kappa^{-2}}\tau} \int_0^\tau ds \int_0^\tau ds'
    \langle \zeta(s) \zeta(s') \rangle e^{\sqrt{1-\kappa^{-2}}(s+s')} \\
    &= \int_0^\tau dy \int_0^\tau dy'
    \langle \zeta(\tau-y) \zeta(\tau-y') \rangle e^{-\sqrt{1-\kappa^{-2}}(y+y')}, \\
    &= 2\varsigma^2 \int_0^\tau dy \int_0^\tau dy'
    \delta(y'-y) e^{-\sqrt{1-\kappa^{-2}}(y+y')},
\end{align*}
where we substituted $y=\tau-s$, $y'=\tau - s'$. Taking the limit $\tau\to\infty$,
we end up with
\begin{align*}
    \langle \varepsilon^2 \rangle 
    &= 2\varsigma^2 \int_0^\infty dy\, e^{-2\sqrt{1-\kappa^{-2}}y} =  \frac{\varsigma^2}{\sqrt{1-\kappa^{-2}}}.
\end{align*}
Plugging this expression back into Eq.~\eqref{eq:two-rsqr}, we obtain
\begin{align*}
    \langle R^2 \rangle = R_0^2 - \frac{1}{4} \varsigma^2.
\end{align*}

\section{Centered dynamics of complex networks}
Eq.~\eqref{eq:network} from the main paper contains a freedom
of re-defining $\theta_i \rightarrow \theta_i +c$ for some constant
$c$ corresponding to a reference angle.
Here, we fix this freedom by introducing the new variables
\begin{align*}
    \delta_i(t) &= \theta_i(t) - \mu(t) \\
    \mu(t) &=  \frac{1}{N}\sum_j \theta_j(t).
\end{align*}
Taking derivatives and plugging them into Eq.~(1), we find that they satisfy
\begin{align}
        \dot\delta_i &= \omega_i - \frac{1}{N}\sum_j \omega_j(t) + \sum_{j} K_{ij} 
        \sin(\delta_i - \delta_j) + \eta_i - \frac{1}{N}\sum_j \eta_j(t)  \label{eq:fixed-dyn}\\
        \dot\mu &= \frac{1}{N}\sum_j \omega_j(t)
        + \frac{1}{N}\sum_j \eta_j(t), \label{eq:mean-dyn}
\end{align}
where we used $\sum_{i,j}K_{ij} \sin(\delta_i-\delta_j) = 0$ due to
antisymmetry.
Equation~\eqref{eq:fixed-dyn} is equivalent to Eq.~\eqref{eq:network} but with
centered inputs, $\omega_i \to \omega_i - (1/N)\sum_j \omega_j(t)$
and $\eta_i \to \eta_i - (1/N)\sum_j \eta_j(t)$.
The order parameter $R^2$ is independent of $\mu$, so it is sufficient
to consider centered dynamics and assume that $\sum_j \omega_j = 0$.
However, we must consider the effect of centering on the stochastic
inputs. Specifically, the centered noisy inputs can be written
using a projection matrix $Q$ as
\begin{align*}
  Q \boldsymbol{\eta} = \left(\mathbb{1} - \frac{1}{N}J\right)\boldsymbol{\eta},
\end{align*}
where $J_{ij}=1$. Similarly, if $\langle \eta_i(t) \eta_j(t') \rangle = C_{ij}
\delta(t-t')$, then the centered correlation matrix is
\begin{align}
  Q \langle \boldsymbol{\eta}(t) \boldsymbol{\eta}^\top(t') \rangle Q^\top = QCQ\, \delta(t-t'). \label{eq:centered-c}
\end{align}
Thus, the centered covariances are constrained to have vanishing
row and column sums, $\sum_j C_{ij} = \sum_i C_{ij} = 0$.
This is also automatically enforced by the Lyapunov equation
constraint in Eq.~\eqref{eq:opt-prob}.

\section{Numerical methods}
\label{sect:numerics}
\subsection{Numerical method for obtaining fixed points}
Fixed points $\bar{\theta}_i$ satisfy
\begin{align*}
    0 = \omega_i + \sum_j K_{ij} \sin(\bar{\theta}_j - \bar{\theta}_i)
\end{align*}
and are found using a Trust-Region method as implemented
in the \texttt{NLsolve.jl} package for the Julia language.
As initial guess we use random numbers drawn from a normal distribution
with mean $0$ and standard deviation $0.01$.

\subsection{Numerical integration of the dynamical equations}
The dynamics in the main paper can be brought into the form of a system of stochastic differential
equations,
\begin{align*}
    d\mathbf{X} = \mathbf{f}(\mathbf{X}, t)\, dt + G\, d\mathbf{W}. 
\end{align*}
Here, $\mathbf{f}(\mathbf{X}, t)$ is the deterministic dynamics and
$d\mathbf{W}$ is a vector of white noise terms. Finally, $G$ is a matrix describing
the correlations between the individual noise terms. For instance, $G=\mathbb{1}$ corresponds
to uncorrelated white noise.
In our work, we prescribe the correlation matrix $C$ between the noise terms,
\begin{align}
    C = \langle G d\mathbf{W} d\mathbf{W}^\top G^\top \rangle = 
     G \underbrace{\langle d\mathbf{W} d\mathbf{W}^\top \rangle}_{=\mathbb{1}} G^\top 
     = G G^\top, \label{eq:c_eq}
\end{align}
where we used the fact that $d\mathbf{W}$ is uncorrelated white noise. 
To construct a matrix $G$ satisfying Eq.~\eqref{eq:c_eq}, we use the
singular value decomposition of $C$,
\begin{align*}
    C = U \Sigma U^\top = U \sqrt{\Sigma} U^\top U \sqrt{\Sigma} U^\top.
\end{align*}
Here, we used that $C$ is symmetric and added in a factor of $U^\top U =
\mathbb{1}$ to end up with a symmetric $G = U \sqrt{\Sigma} U^\top$ (this specific
choice has no bearing on the results).

To efficiently calculate the long-time averaged order parameter without
the need to store the entire time series,
we note that any time average $y(t) = \langle f(t) \rangle = \frac{1}{t} \int_0^t f(t') dt'$ satisfies the differential equation
\begin{align}
    y'(t) &= \frac{f(t) - y(t)}{t}, \label{eq:ode-avg}
\end{align}
with the initial condition $y(0) = f(0)$ and $y'(0) = \frac{1}{2} f'(0)$.
Equation~\eqref{eq:ode-avg} with $f(t) = R^2(t)$ is solved concurrently with 
the original SDE and produces the averaged order parameter.

The complete system of SDEs is numerically solved using the
\texttt{DifferentialEquations.jl}
package~\cite{Rackauckas2017} in the Julia language with an Euler-Maruyama scheme and time-step
$\Delta t=0.01$.

\subsection{Numerical optimization}
Numerical optimization of the problem given by Eq.~\eqref{eq:opt-prob}
is done by implementing it in the
domain-specific language of the \texttt{Convex.jl} package~\cite{convexjl} for the
Julia language.
The problem is then solved using the COSMO algorithm~\cite{Garstka2019}
with convergence tolerances $\varepsilon_\text{rel} = \varepsilon_\text{abs} = 10^{-7}$.

\section{Derivation of the Lyapunov equation constraint}
\label{sect:lyap}
Here, we compute the variance of fluctuations directly in the
Langevin formalism.
We are interested in the long-term limit where any initial transients have decayed,
and the system has relaxed to a stationary distribution.

We consider the linearized first-order system from the main paper,
\begin{align*}
  \dot{\boldsymbol{\varepsilon}} = L\boldsymbol{\varepsilon} +  \boldsymbol{\eta}(t),
\end{align*}
where $\langle \boldsymbol{\eta} \rangle = \vec0$, $\langle \boldsymbol{\eta}(t) \boldsymbol{\eta}(t')^\top \rangle = C\, \delta(t - t')$
is white noise input in time with centered correlation matrix $C$ and $L$
is the weighted graph Laplacian with components $L_{ij} = K_{ij} \cos(\bar\theta_j - \bar\theta_i) - \delta_{ij} \sum_n K_{in} \cos(\bar\theta_n - \bar\theta_i) $.
The solution to this system can be expressed as
\begin{align*}
  \boldsymbol{\varepsilon}(t) = \exp(L t) \boldsymbol{\varepsilon}_0 + \int_0^t \exp(L(t- s)) \boldsymbol{\eta}(s) \,ds.
\end{align*}
As long as the network graph is connected, the Laplacian
has a single vector in its nullspace, 
the vector of all 1s $\mathbf{1}=(1,1,\dots,1)^\top$,
and is otherwise negative definite. The centered
dynamics introduced in the previous
section are equivalent to $\mathbf{1}^\top\boldsymbol{\varepsilon}$,
such that all solutions to the linear differential equation live in the space
orthogonal to the Laplacian's nullspace.
Thus, the homogeneous solution
$\exp(L t) \boldsymbol{\varepsilon}_0 = \sum_{j=2}^N e^{\lambda_j t}
(\mathbf{u}_j^\top \boldsymbol{\varepsilon}_0) \mathbf{u}_j$,
where $\lambda_j < 0$ are the nonzero eigenvalues and $\mathbf{u}_j$ the corresponding
eigenvectors of $L$, decays for large times, and we can focus on the particular solution.

\noindent We want to compute the matrix of equal-time covariances in the long-time limit,
\begin{align}
  \langle \boldsymbol{\varepsilon}(t) \boldsymbol{\varepsilon}(t)^\top \rangle &= \int_0^t d{s} \int_0^{t} d{s}' \exp(L(t- {s})) \langle \boldsymbol{\eta}({s}) \boldsymbol{\eta}({s}')^\top \rangle  \exp(L(t- {s}'))
   \nonumber \\
  &= \int_0^t d{s} \int_0^{t} d{s}' \exp(L(t-s)) C   \exp(L(t-s')) \, \delta({s} - {s}') \nonumber \\
  &= \int_0^t dy \int_0^{t} dy' \exp(Ly) C   \exp(Ly') \, \delta( y'-y).
  \label{eq:correlation}
\end{align}
We substituted $y= t - {s}, y' = t - {s}'$, and used the fact
that
  $\langle \boldsymbol{\eta}(t) \boldsymbol{\eta}^\top(t') \rangle =  C\, \delta(t-t').$
We now take the limit $t\to\infty$. Integrating
over the $\delta$-function we obtain
\begin{align}
  \langle \boldsymbol{\varepsilon} \boldsymbol{\varepsilon}^\top \rangle
  &= \int_0^\infty dy \exp(Ly) C  \exp(Ly) \nonumber
\end{align}
This matrix-valued integral cannot be evaluated directly, but we can integrate by parts
to obtain
\begin{align}
  \langle \boldsymbol{\varepsilon} \boldsymbol{\varepsilon}^\top \rangle = E 
  &= \int_0^\infty d{y} \exp(L {y}) C \exp(L {y}) \nonumber \\
  &= \left[\exp(L{y})C\exp(L{y}) \right]_0^\infty L^\dagger  - L\int_0^\infty d{y} \exp(L {y}) C \exp(L {y}) L^\dagger \nonumber \\
  &= -CL^\dagger - L E L^\dagger \nonumber \\
  \Rightarrow LE + EL &= -C.
  \label{eq:lyapunov}
\end{align}
Equation~\eqref{eq:lyapunov} is the continuous Lyapunov equation,
which frequently occurs in control and stability theory~\cite{Gajic2008}. For instance, a linear time-invariant system
given by the matrix-valued ODE $A\dot{\mathbf{x}} = \mathbf{x}$
is globally asymptotically stable if the Lyapunov equation $A^\top P + PA = -Q$ can be solved for any positive-definite $Q$.
Here, the dagger represents the Moore-Penrose pseudo-inverse, which
is used because the nullspace of $L$ and $C$ is given by the vector
$\mathbf{1}$  of all ones.
The solution $E$ must also have $E\mathbf{1} = 0$ in the centered frame,
such that Eq.~\eqref{eq:lyapunov} captures it fully.

\section{The order-parameter Hessian near the synchronous state}
The Hessian of the squared order parameter for a network of $N$ oscillators 
near the fixed point $\bar{\theta}_j$ is the $N\times N$
matrix with elements
\begin{align*}
    H_{ij} = \frac{2}{N^2} \left( \cos(\bar{\theta}_i - \bar{\theta}_j)
    - \delta_{ij} \sum_{n=1}^N \cos(\bar{\theta}_i - \bar{\theta}_n) \right).
\end{align*}

\subsection{The synchronous state}
At the synchronous state $\bar{\theta}_i = 0$, the Hessian has the form
\begin{align*}
    H_{ij}^{(0)} = \begin{cases}
    1 - N & i= j \\
    1 & \text{otherwise}.
    \end{cases}
\end{align*}
It is easy to see that the vector of all ones lies in the nullspace, $H\mathbf{1} = 0$.
Similarly, the set of $N-1$ vectors $(1,-1,0,\dots, 0), (1,0,-1,0,\dots,0), \dots,
(1,0,\dots,0,-1)$ provides a basis of the remaining eigenspace corresponding to
the eigenvalue $-2/N$. Thus, the Hessian is negative semi-definite.
We can approximate the eigenvalues for not perfectly synchronized states using
perturbation theory. We first note that $H\mathbf{1} = 0$ is always true due to the
structure of the matrix.
We then expand the cosines to find the first order correction of $H = H^{(0)} + H^{(1)} + \dots$
as
\begin{align*}
    H_{ij}^{(1)} = -\frac{1}{N^2} \left( (\bar{\theta}_i - \bar{\theta}_j)^2
    - \delta_{ij} \sum_{n=1}^N (\bar{\theta}_i - \bar{\theta}_n)^2 \right)
\end{align*}
It is easy to check that this matrix is negative semi-definite,
\begin{align*}
    \sum_{i,j} x_i H_{ij}^{(1)} x_j = -\frac{1}{2N} \sum_{i,j} (\bar{\theta}_i - \bar{\theta}_j)^2
    (x_i - x_j)^2 \leq 0.
\end{align*}
Thus, the eigenspace of $H^{(0)}$ corresponding to the eigenvalue
$-2/N$ is perturbed to first order to $-2/N + \lambda_i$, where $\lambda_i$
are the eigenvalues of $H^{(1)}$. We conclude that close to the synchronous state, 
the order parameter Hessian $H$ is negative semi-definite.

\subsection{Twisted states in periodic chains}
Here, we analyze periodic chains of $N$ oscillators near stable ``twisted states.''
Following Ref.~\cite{Wiley2006}, twisted states are defined 
by $\bar{\theta}_j = 2\pi \frac{q}{N} j$, where $q=1,2,\dots,N-1$ is the
winding number.

\begin{figure}
    \centering
    \includegraphics[width=.7\textwidth]{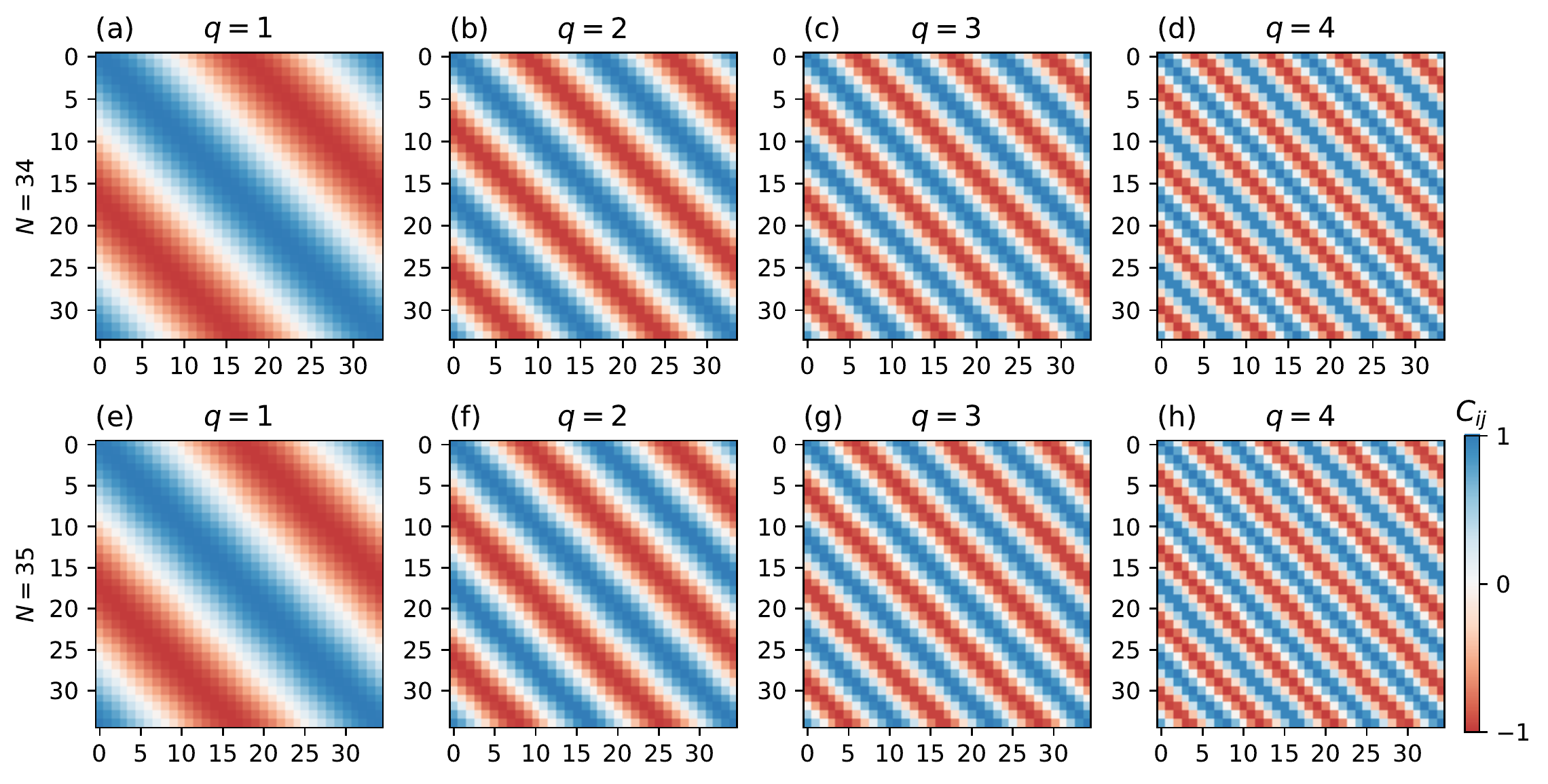}
    \caption{Optimal noise covariances for twisted states with $q=1,2,3,4$ in periodic chains. (a--d) Chains of
    length $N=34$. (e--h) Chains of length $N=35$. In both cases, the optimal noise pattern is oscillatory
    with a wavelength that decreases with the winding number $q$.}
    \label{fig:twisted_C}
\end{figure}

The Hessian at a twisted state is
\begin{align*}
    H^{(0)}_{ij} = \frac{2}{N^2} \cos\left( 2\pi \frac{q}{N} (i - j) \right),
\end{align*}
because the sum $\sum_n\cos\left( 2\pi \frac{q}{N} (i - n) \right) = 0$. This matrix
is circulant, so the eigenvalues are given by
\begin{align*}
    \lambda_j &= \sum_{k=1}^N \frac{2}{N^2} \cos\left( 2\pi \frac{q}{N} (k - 1) \right)
    e^{2\pi i \frac{k-1}{N}j}, \qquad j=0,\dots, N-1 \\
    &= \frac{1}{N^2} \sum_{k=0}^{N-1} \left[ \left( e^{2\pi i \frac{j+q}{N}} \right)^k 
    + \left( e^{2\pi i \frac{j-q}{N}} \right)^k \right] \\
    &= \begin{cases}
    \frac{1}{N} & j = q \\
    \frac{1}{N} & j = N-q \\
    0 & \text{ otherwise}.
    \end{cases}
\end{align*}
Thus, the Hessian at a twisted state fixed point is positive semi-definite.
Expanding near a twisted state, $\bar{\theta}_j = 2\pi qj/N + \epsilon_j$ the lowest order correction
to the Hessian can be written as
\begin{align*}
    H^{(1)}_{ij} = -\sin \left( 2\pi \frac{q}{N} (i - j) \right)(\epsilon_i - \epsilon_j)
    + \delta_{ij} \sum_n \sin \left( 2\pi \frac{q}{N} (i - n) \right)(\epsilon_i - \epsilon_n).
\end{align*}
This matrix is indefinite in general, such that the full Hessian eigenvalues close to twisted states
are generally indefinite. Numerically it can be seen that near a twisted state, the
0-eigenvalues of $H^{(0)}$ tend to split into positive and negative pairs of eigenvalues.

\section{Optimal noise for twisted states in periodic chains}
Here, we analyze periodic chains of $N$ oscillators near stable ``twisted states.''
Following Ref.~\cite{Wiley2006}, we set $\omega_i = 0$ and choose twisted initial
conditions, $\bar{\theta}_j = 2\pi \frac{q}{N} j$, where $q=1,2,\dots,N-1$ is the
winding number. Such twisted states can be thought of as maximally asynchronous
because the order parameter is
\begin{align*}
    R = \frac{1}{N} \sum_{j=1}^N e^{2\pi i \frac{q}{N} j} = 0.
\end{align*}
Thus, near stable twisted states we expect any amount of noise to increase the order parameter. Specifically, we find that our method as outlined in the main paper still works and provides optimal noise patterns
which improve $\langle R^2 \rangle$ compared to uncorrelated noise and to the no-noise case.

\begin{figure}
    \centering
    \includegraphics[width=.53\textwidth]{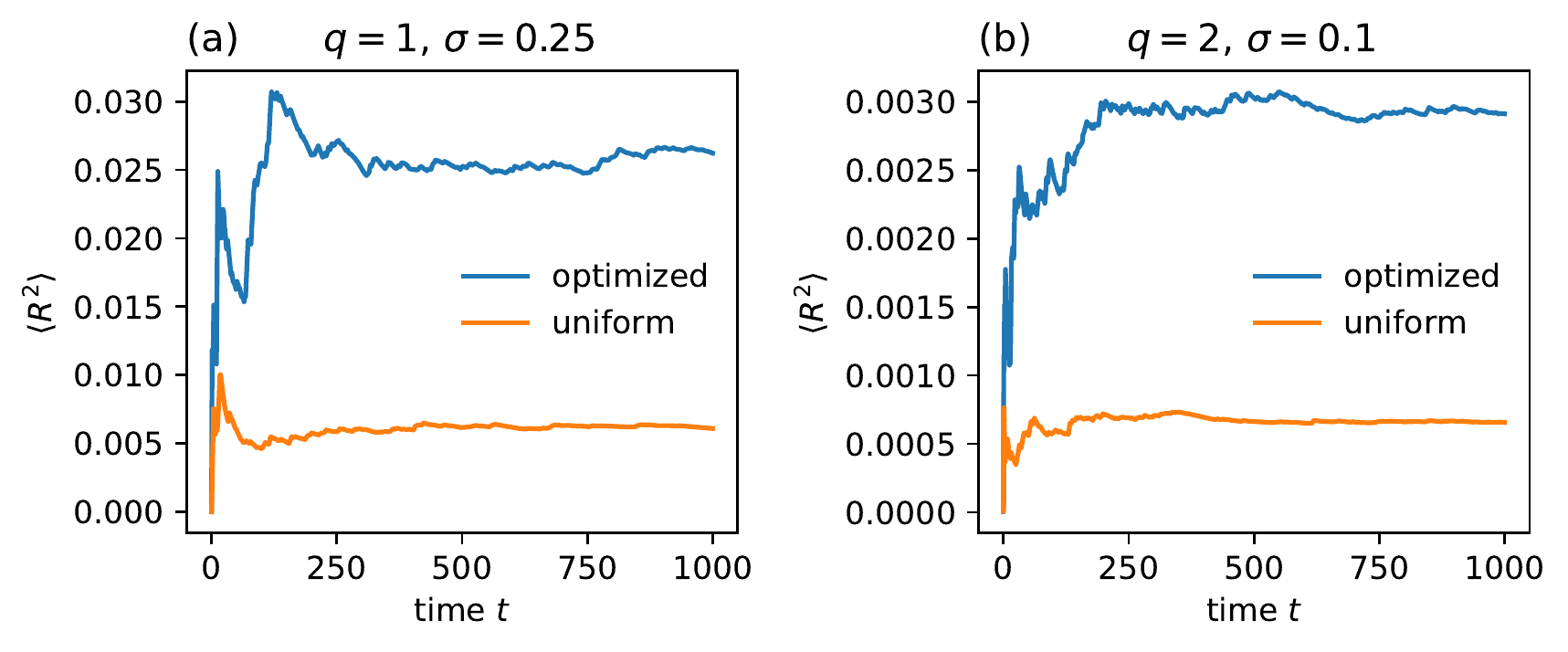}
    \caption{Time series of the time averaged order parameter
    $\langle R^2 \rangle(t) = (1/t)\int_0^t R^2(t') dt'$ in an $N=10$ periodic
    chain near twisted states with winding numbers (a) $q=1$ and (b) $q=2$. 
    For panel (a), $\sigma=0.25$ and for panel (b), $\sigma=0.1$.}
    \label{fig:twisted_C_ts}
\end{figure}

The optimal noise covariances are no longer anticorrelated between neighboring oscillators but oscillate
spatially with a wavelength that decreases with increasing winding number $q$ (Figure~\ref{fig:twisted_C}).

Direct numerical simulations show that optimal noise patterns do improve
synchrony as compared to uniform noise (Figure~\ref{fig:twisted_C_ts}).
Because for higher winding numbers $q$ the basins of attraction of the
twisted states become smaller, it becomes steadily more difficult to remain
near the twisted state in a noisy system. For instance, in the $N=10$
chain from Figure~\ref{fig:twisted_C_ts}, states with $q>2$ were not stable
for any amount of noise $\sigma>10^{-6}$.

\section{Optimal noise in periodic square grids}
\begin{figure*}
    \centering
    \includegraphics[width=.8\textwidth]{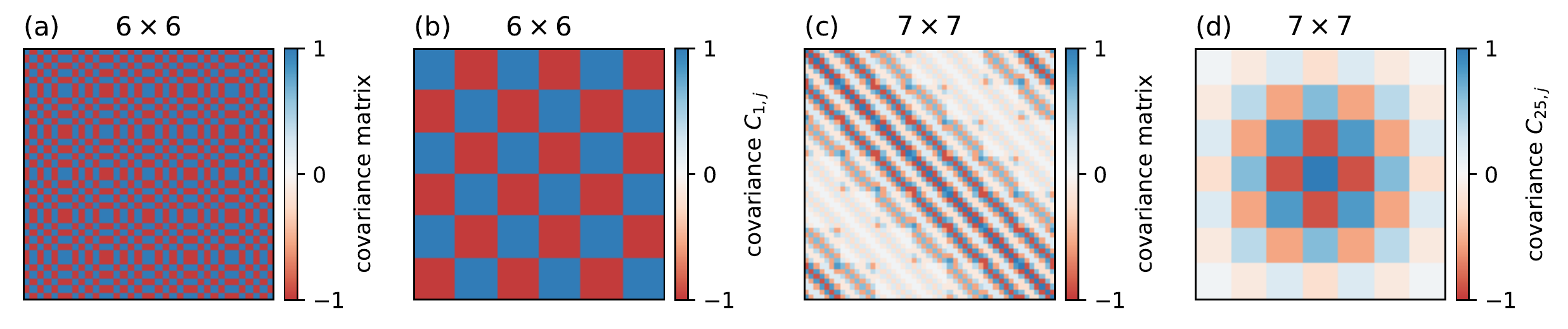}
    \caption{Optimal noise patterns and frustrated covariances in periodic square
    grids. 
    (a) Optimal covariance matrix of a $6\times 6$ periodic
    square grid of oscillators. (b) Covariance pattern for the same grid as in panel (a)
    with respect to the top left oscillator, all neighbors are anti-correlated.
    Each matrix entry corresponds to one
    oscillator in the grid with neighbors as shown.
    (c) Optimal covariance matrix of a frustrated $7\times 7$ periodic
    square grid of oscillators. Because the graph has an odd number of oscillators in each direction, not all
    neighbors can receive anti-correlated noise.
    (d) Covariance pattern for the same grid as in panel (c)
    with respect to the center oscillator. Covariances decay with distance,
    similar to Fig.~\ref{fig:chain-frust} (d,f).}
    \label{fig:grid-frust}
\end{figure*}
Periodic grids show similar behavior as oscillator chains.
In a periodic even square grid, the optimal noise is perfectly anti-correlated between
neighboring nodes [Fig.~\ref{fig:grid-frust} (a,b)],
whereas for odd square grids, correlations show a characteristic
decay similar to odd chains due to the fact that not all neighbors
can receive perfectly anti-correlated inputs [Fig.~\ref{fig:grid-frust} (c,d)].

\section{Parametric dependence of the optimal covariance in periodic chains and optimal noise transition}
In the following section we study the dependence of the optimal noise covariance matrix on the
parameters of the system, the coupling constant $K$ and the distribution of natural frequencies
$\omega_i$. For simplicity, we focus on periodic oscillator chains and on Gaussian distributed
natural frequencies with zero mean where we vary the width of the distribution.
For couplings $K$ close to the synchronization transition, a new transition
in the optimal noise covariances is observed from small-scale to large-scale correlations.
The width of the transition region depends on the width of the distribution of natural
frequencies.
The optimal noise covariance matrix obtained close to the synchronization transition in the
phase-locked regime still provides a significant improvement
in the order parameter even in the phase-drift regime, for a large range
of parameters even beyond the noise-free order parameter.

\subsection{Dependence on the coupling constant and noise strength}
\begin{figure}
    \centering
    \includegraphics[width=0.99\textwidth]{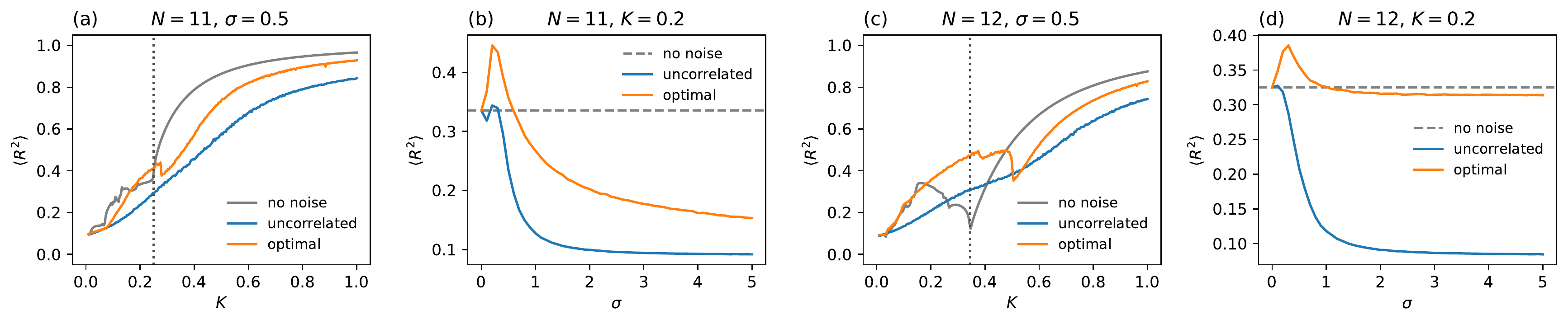}
    \caption{Dependence of the long-time averaged order parameter $\langle R^2 \rangle$ on the parameters $K$ and $\sigma$
    in periodic chains of (a,b) $N=11$ and (c,d) $N=12$. Panels (a,c) show the order parameter without noise
    (from direct numerical simulations and from theory by numerically calculating the fixed point),
    with optimal noise, and with uncorrelated noise of equal total variance. The optimal noise covariance matrix $C$ was calculated for
    each value of $K$ independently. In the phase-drift regime, the matrix $C$ corresponding to the closest
    value of $K$ where a fixed point existed was used. The vertical dotted lines
    correspond to the critical values $K_\text{crit}$ where the synchronization transition occurred.
    Numerical simulation were performed from initial conditions either at the
    numerically obtained noise-free fixed point (in the phase-locked regime) 
    or from zero initial conditions (in the phase-drift regime)
    up to a time $t=200000$.}
    \label{fig:chain-numerics}
\end{figure}
Here we study the dependence of the optimal covariance matrix on the coupling $K$ in periodic chains.
We fixed chains of length $N=11,12$ as in the
main paper with random but fixed distribution of natural frequencies $\{\omega_i\}$ 
where the standard deviation was $\operatorname{std}(\{\omega_i\}) = 1/N$.
We varied the uniform coupling constant $K$ between $K=0$ and $K=1$.

We numerically integrated the stochastic equations of motion from initial conditions at the
steady state if one existed (computed as in Section~\ref{sect:numerics})
or from zero initial conditions (if no steady state existed)
for (i) no noise, (ii) uncorrelated noise, and
(iii) optimal noise.
When a steady state existed at a given $K$, the optimal noise was obtained using the same method
as described in the main paper. When no steady state existed at a given $K$, the optimal noise from
the closest $K$ where a steady state existed was used. The noise variance was fixed to $\sigma=0.5$
and the equations of motion were integrated until $t=200000$.

\begin{figure}
    \centering
    \includegraphics[width=0.7\textwidth]{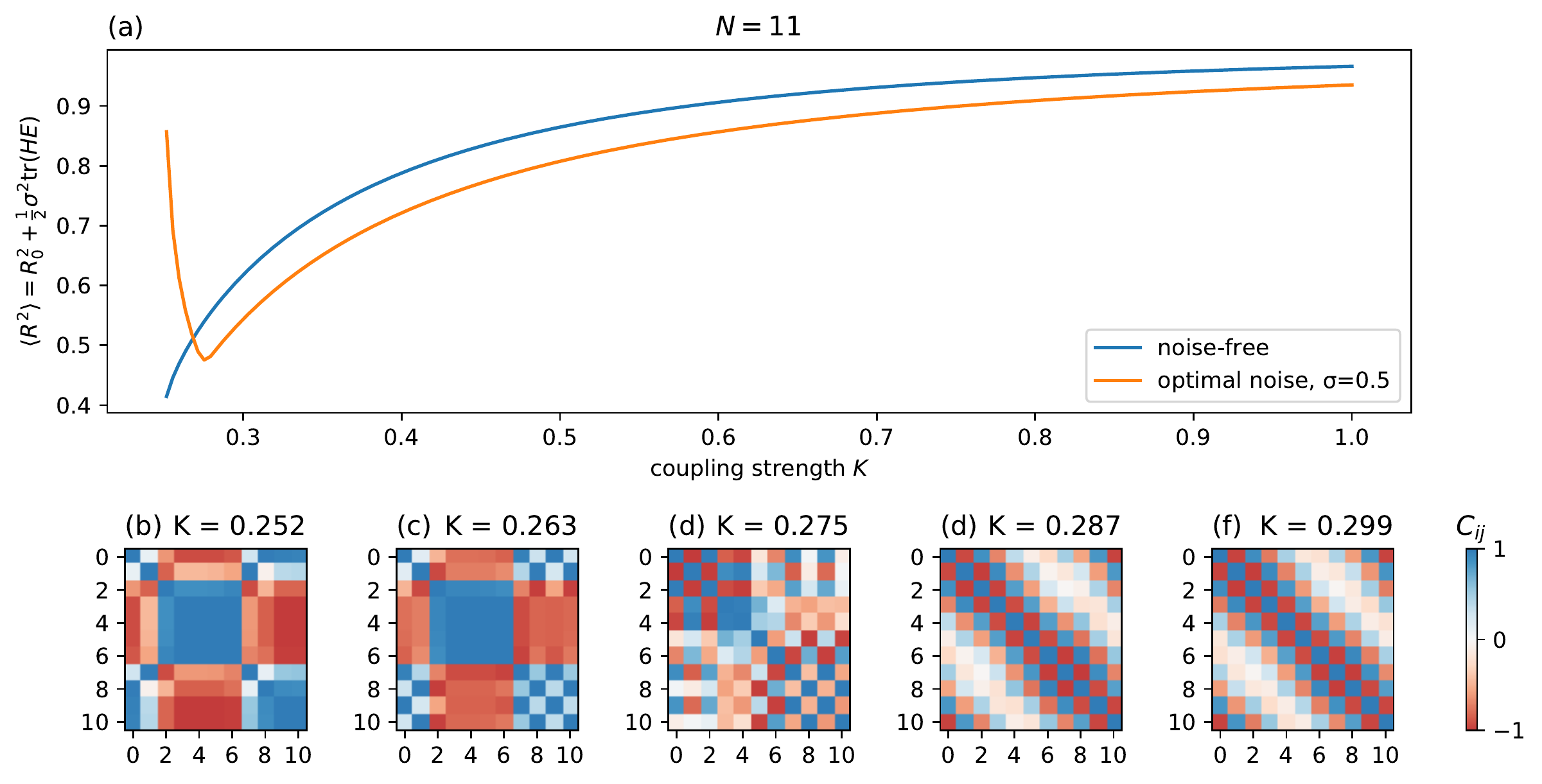}
    \caption{Optimal noise covariances near the synchronization transition in an $N=11$ chain. (a)
    Noise-free and approximate optimal noise order parameters as a function of the coupling strength
    $K$.  The optimal noise curve crosses the no-noise curve near the synchronization transition. The curves correspond
    to those in Fig.~\ref{fig:chain-numerics}~(a).
    (b--e) Optimal covariance matrices near the synchronization transition, exhibiting
    a transition from small scale to large scale correlations.}
    \label{fig:optimal-Kdep-11}
\end{figure}

\begin{figure}
    \centering
    \includegraphics[width=0.7\textwidth]{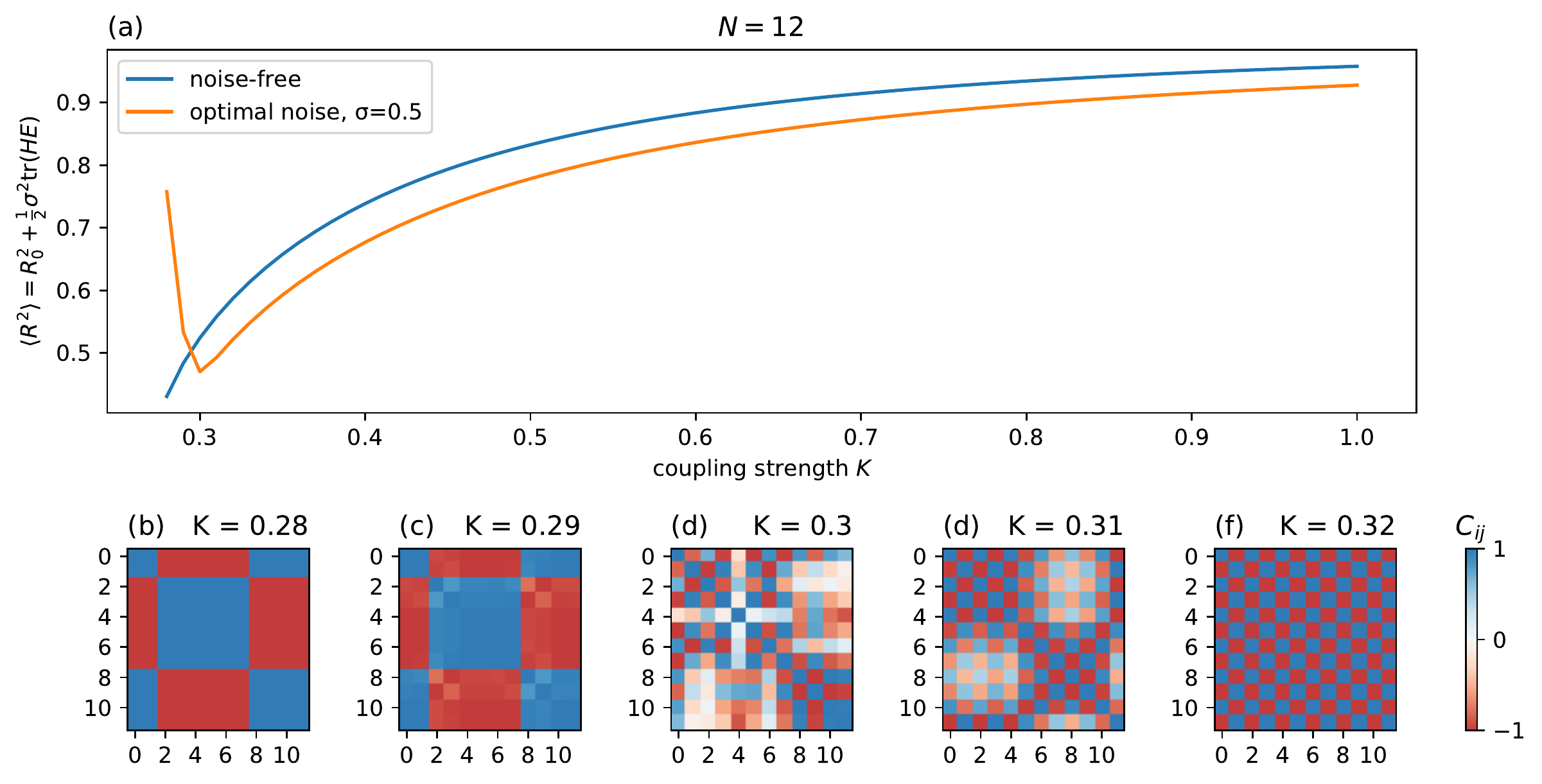}
    \caption{Optimal noise covariances near the synchronization transition in an $N=12$ chain. (a)
    Noise-free and approximate optimal noise order parameters as a function of the coupling strength
    $K$. The optimal noise curve crosses the no-noise curve near the synchronization transition. 
    The curves correspond to those in Fig.~\ref{fig:chain-numerics}~(c).
    (b--e) Optimal covariance matrices near the synchronization transition, exhibiting
    a transition from small scale to large scale correlations.}
    \label{fig:optimal-Kdep-12}
\end{figure}

Generally, we found that in the steady-state regime the behavior was as expected from the main paper (Figure~\ref{fig:chain-numerics}). Specifically, it is possible that the choice
of natural frequencies $\omega_i$ is such that close to the synchronization transition, the optimal noise
also undergoes a transition and becomes significantly better at synchronizing the network than even 
the no-noise case [Fig.~\ref{fig:chain-numerics} (a,c)].
The transition of the optimal noise covariance matrix depends on the network
topology and proceeds towards more large-scale correlations as the
synchronization transition is approached (Figures~\ref{fig:optimal-Kdep-11}, 
\ref{fig:optimal-Kdep-12}).

\subsection{Dependence on the frequency distribution}
To study the dependence on the frequency distribution, we again consider periodic chains of
various sizes. For simplicity, we consider only Gaussian distributed natural frequencies $\omega_i$, but
we vary the the standard deviation, $\operatorname{std}(\omega)$.
We choose periodic oscillator chains of sizes $N=10,12,15$ and perform numerical 
optimizations of the noise covariances. For each oscillator chain, the distribution
of $\omega_i$'s is fixed and then the initial distribution is uniformly rescaled to
obtain different standard deviations.

\begin{figure}
    \centering
    \includegraphics[width=.8\textwidth]{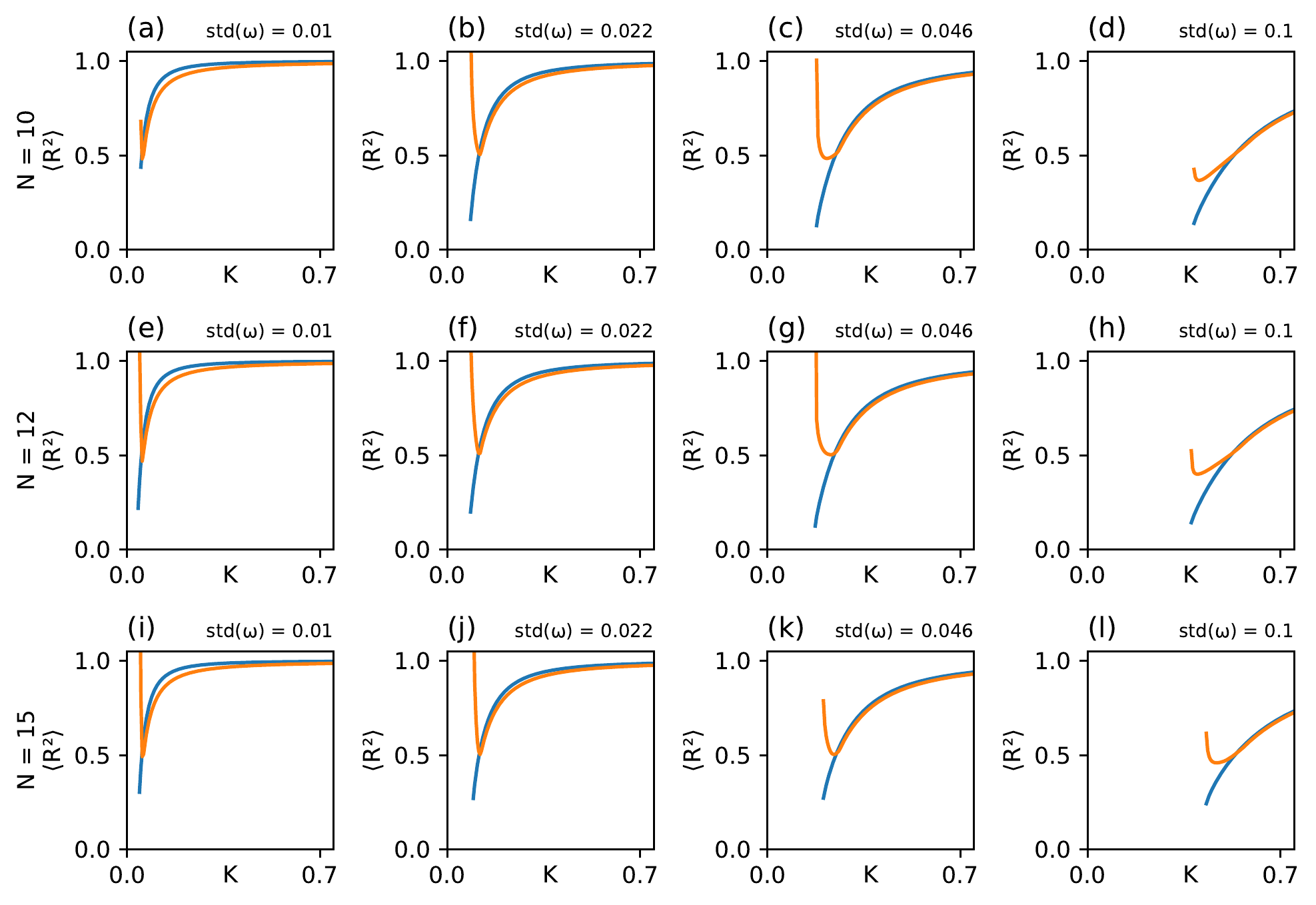}
    \caption{Dependence of the optimal noise transition on the frequency distribution.
    (a--c) $N=10$ periodic chain, (e--f) $N=12$ periodic chain, (i--l) $N=15$ periodic
    chain. Shown are the steady-state order parameters $R_0^2$ without noise (blue lines) and 
    the approximations $\langle R^2 \rangle = R_0^2 + (\sigma^2/2) 
    \operatorname{tr}(HE)$ with $\sigma=0.25$ (orange lines).
    The synchronization transition where steady states cease to exist occurs at the
    value of $K$ where both lines end.
    }
    \label{fig:omegas}
\end{figure}

In all cases, we find the same transition if the optimal noise covariance near the
synchronization transition of the original system. However, the transition
happens over a smaller range of $K$ as the standard deviation $\operatorname{std}(\omega)$
is decreased (Figure~\ref{fig:omegas}).

\section{Second-order Power grid models}
Power grids are often modeled using the swing equation (a
second-order Kuramoto model), which includes the effects of
mechanical inertia in the grid~\cite{Nardelli2014}. The equations of
motion for $N$ nodes are
\begin{align}
    \dot{\theta}_i &= \omega_i \nonumber \\
    \dot{\omega}_i &= -\alpha\, \omega_i + \sum_{j=1}^N
    K_{ij} \sin(\theta_j - \theta_i) + P_i + \eta_i. \label{eq:swing}
\end{align}
Here, $\alpha$ is a damping constant,
$K_{ij}$ are coupling constants related to physical properties
of the grid power lines, and the constants $P_i$ are related to net power flows
(generation or consumption) at each node $i$. The $\eta_i$ are noise terms
with $\langle \eta_i \rangle = 0$ and $\langle \eta_i(t) \eta_j(t') \rangle = C_{ij} \delta(t-t')$. We will also assume
that the net powers are balanced, $\sum_j P_j = 0$.

Just like in the main manuscript, we can find a noise-free steady state
by solving
\begin{align*}
    0 &=  \sum_{j=1}^N
    K_{ij} \sin(\bar{\theta}_j - \bar{\theta}_i) + P_i,
\end{align*}
which has the same mathematical form as the equivalent equation in the Kuramoto model.
Expanding near this state as $\theta_i(t) = \bar{\theta}_i + \varepsilon_i(t)$ for small
$\varepsilon_i$ we obtain the linearized equations of motion
\begin{align*}
    \underbrace{\begin{pmatrix}
    \dot{\bm{\varepsilon}} \\
    \dot{\bm{\omega}}
    \end{pmatrix}}_{=\dot{\mathbf{y}}}
    =
    \underbrace{\begin{pmatrix}
    0 & \mathbb{1} \\
    L & -\alpha \mathbb{1}
    \end{pmatrix}}_{= M}
    \underbrace{\begin{pmatrix}
    \bm{\varepsilon} \\
    \bm{\omega}
    \end{pmatrix}}_{=\mathbf{y}}
    + 
    \begin{pmatrix}
    0 \\
    \bm{\eta}(t)
    \end{pmatrix}.
\end{align*}
Here, we again used the Laplacian with components $L_{ij} = K_{ij} \cos(\bar\theta_j - \bar\theta_i) - \delta_{ij} \sum_n K_{in} \cos(\bar\theta_n - \bar\theta_i)$.
For this linearized system, we can go through the same derivation as
in Section~\ref{sect:lyap} to show that the long-time covariance matrix $F = \langle\mathbf{y} \mathbf{y}^\top\rangle$ satisfies another Lyapunov equation,
\begin{align*}
    M F + F M^\top = -\begin{pmatrix}
    0 & 0 \\
    0 & C
    \end{pmatrix} = -\hat{C}.
\end{align*}
Note the transpose `${}^\top$' which is due to the fact that $M$ is not symmetric.

The synchrony of a power grid can be quantified by the same type of long-time averaged
order parameter as in the Kuramoto model,
\begin{align*}
    \langle R^2 \rangle &= R_0^2 + \frac{1}{2} \langle \bm{\varepsilon}^\top H \bm{\varepsilon} \rangle + \dots \\
        &= R_0^2 + \frac{1}{2} \operatorname{tr}(H E) + \dots
\end{align*}
Here, $H$ is the same Hessian as in the main paper and the matrix $E$ of covariances
between the $\bm{\varepsilon}$'s is the top left $N\times N$ block of
\begin{align*}
    F = \begin{pmatrix}
    E & * \\
    * & *
    \end{pmatrix}.
\end{align*}
Armed with this model, we can formulate the optimal noise 
problem for power grids as
\begin{align}
    \max_{C,F} &\quad  \operatorname{tr}\left(
    H E\right) \label{eq:swing-opt} \\
\text{such that }&\quad MF + FM^\top = -\hat{C} \nonumber \\
    &\quad  C \succeq 0. \nonumber
\end{align}

\begin{figure}
    \centering
    \includegraphics[width=0.55\textwidth]{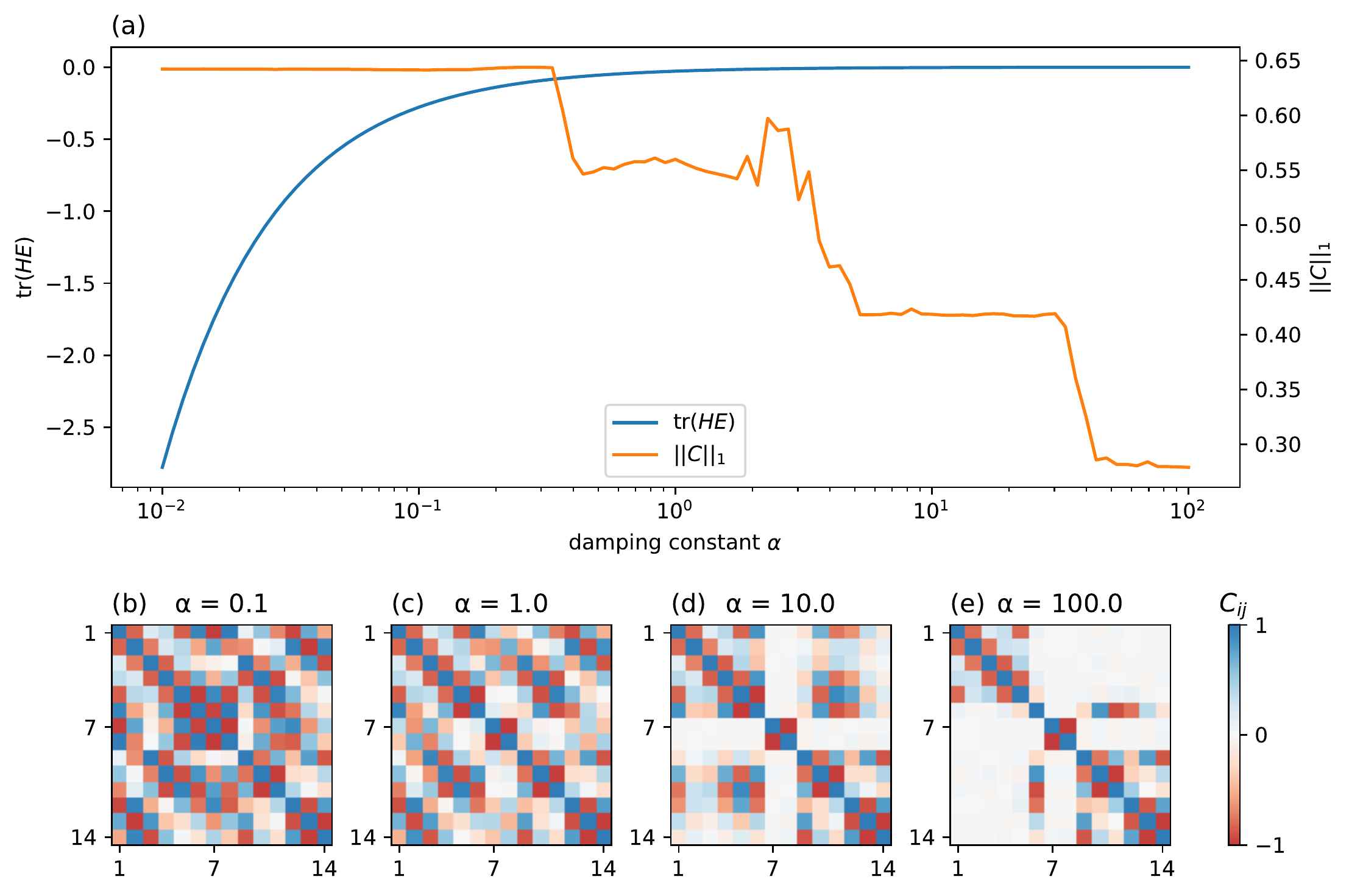}
    \caption{Optimal noise covariance matrices $C$ in the IEEE 14-bus test grid
    with the swing equation model Eq.~\eqref{eq:swing-opt}. (a) We plot the optimal value of the objective function $\operatorname{tr}\left(H E\right)$ against the
    damping parameter $\alpha$, and we quantify the structure of the optimal covariance matrix
    using its Frobenius 1-norm $\|C\|_1 = \sum_{i,j}|C_{ij}|$, revealing a
    number of transitions (b--e).}
    \label{fig:ieee14_swing}
\end{figure}

We numerically optimized the IEEE 14-bus test case near its
synchronous steady state for various
values of $\alpha$ (Fig.~\ref{fig:ieee14_swing}). 
For large $\alpha \gg 1$, the model approaches the Kuramoto case,
and the optimal covariances are clustered.
Decreasing $\alpha$, and thus increasing the relative importance
of inertia, the optimal covariances become less clustered
by going through a number of step-like transitions in the
covariance structure. At the same time, the optimal objective
$\operatorname{tr}(HE)$ decreases: in strongly inertia-dominated
networks, noise optimization is less effective at increasing synchrony.
However, the basic anti-correlated structure of optimal
covariances remains.

\end{document}